\documentclass[11pt]{article}  

\usepackage{amsmath,amsthm,latexsym,amssymb,amsfonts,epsfig,braket,mathtools,yhmath}
\usepackage{xcolor}
\usepackage{cleveref}
\usepackage{dsfont}

\allowdisplaybreaks

\newcommand{\be}{\begin{equation}}
\newcommand{\ee}{\end{equation}}
\newcommand{\ba}{\begin{eqnarray}}
\newcommand{\ea}{\end{eqnarray}}

\title{{\sf Hamiltonian renormalisation IX. $U(1)^3$ quantum gravity}}
\author{
{\sf A. M. Rodriguez Zarate}$^1$\thanks{{\sf 
melissa.rodriguez@gravity.fau.de}},
{\sf T. Thiemann}$^1$\thanks{{\sf 
thomas.thiemann@gravity.fau.de}}\\
\\
{\sf $^1$ Inst. for Theor. Phys. III, FAU Erlangen -- N\"urnberg,}\\
{\sf Staudtstr. 7, 91058 Erlangen, Germany}\\
}
\date{{\small\sf \today}}

\makeatletter
\@addtoreset{equation}{section}
\makeatother

\begin{document} 

\maketitle

{\sf

\begin{abstract}
In previous works in this series we focussed on Hamiltonian renormalisation of free 
field theories in all spacetime dimensions or interacting theories in spacetime dimensions 
lower than four. In this paper we address the Hamiltonian renormalisation of the $U(1)^3$
model for Euclidian general relativity in four spacetime dimensions which is self-interacting.

The Hamiltonian flow needs as an input a choice of $^\ast-$algebra and corresponding 
representation thereof or state on it at each resolution scale. If one uses as input algebras
and states in analogy to those used in the recent exact solutions of this model,
then one finds that the flow finds as fixed point those exact solution theories .     
\end{abstract}

\section{Introduction}
\label{s1}

Constructing interacting quantum field theories (QFTs) rigorously in four and higher spacetime 
dimensions remains one of the most difficult challenges in theoretical and mathematical 
physics \cite{a}. The difficulties come from the fact that quantum fields are 
operator valued distributions which means that products thereof as they appear 
typically in Hamiltonians are a priori ill-defined, being plagued by both short distance 
(UV) and large distance (IR) divergences. In the constructive QFT (CQFT) approach \cite{b} 
one tames both types of divergences by introducing both UV ($M$) and IR cut-offs ($R$) to the effect 
that only a finite number of degrees survive at finite $M,R$. For instance, $R$ could be 
a compactification radius and $M$ a lattice spacing. Then at finite $M,R$ one is in the safe realm 
of quantum mechanics. The problem is then how to remove the cut-offs. Usually one removes 
first $M$ (continuum limit) and then $R$ (thermodynamic limit). In this process the parameters 
(coupling constants) are taken to be cut-off dependent and they are tuned or renormalised in such a way 
that the limiting theory is well-defined when possible.  

Non-perturbative renormalisation in CQFT (not to be confused with renormalisation 
in the perturbative approach to QFT) has a long tradition \cite{c} and comes 
in both the functional integral language and the Hamiltonian language (see e.g. \cite{d} and 
references therein). Focussing on UV cut-off removal, we consider quantum mechanical
systems labelled by the cut-off $M$. If these quantum mechanical systems all descend form 
a well-defined continuum theory, then in the functional integral approach one obtains 
the theory at resolution $M$ by integrating out all degrees of freedom referring to higher 
resolution while in the Hamiltonian approach one projects those out. This in particular implies 
that if one takes the quantum mechanical theory at resolution $M'$ and 
integrates or projects out the degrees of freedom at resolutions between 
$M<M'$ and $M'$ one obtains the quantum mechanical theory at resolution $M$. 
Vice versa, when this necessary set of {\it consistency conditions} is met, this typically 
also is sufficient to define a continuum theory.  
 
Now the 
family of theories that one starts with are constructed making various choices such 
as representations, factor orderings, discretisation errors etc. and the aforementioned 
consistency conditions are generically violated. However, one can define  
a sequence of such quantum mechanical theory families by defining a new theory at resolution 
$M$ by integrating/projecting out the degrees of freedom between $M$ and $M'(M)$ of the old 
theory at resolution $M'(M)>M$ where $M':\; M\mapsto M'(M)$ is a fixed function on the set of 
resolution scales. Such a process is called a block spin transformation or coarse graining 
operation which typically leads to a renormalisation of the coupling constants. At a fixed point 
of this renormalisation flow of theories the consistency condition is enforced by construction
and therefore fixed points qualify as continuum theories.

In previous parts of this series we have considered a Hamiltonian projection scheme
\cite{LLT1,d} which is motivated by the functional integral approach via 
Osterwalder-Schrader reconstruction. It was then applied to free QFT 
in Minkowski space \cite{LLT2, LLT3, LLT4, LT, TT} in any dimension and parametrised 
QFT \cite{TZ} in 2d which shares some features with the free bosonic string. 
More recently we applied it \cite{RZ-T} to the interacting scalar $P(\Phi)_2$ theory \cite{e}
in 2d and finite volume. In all those cases the fixed point of the flow could be computed and was shown 
to coincide with the known continuum theory.

In the present paper we consider the U(1)$^3$ toy model for Euclidian signature quantum 
gravity in four dimensions \cite{f}. It can be considered as a weak (Newton constant) 
coupling limit of actual Euclidian signature gravity. The model is simpler than the actual 
theory in the sense that the actual non-Abelian gauge group SU(2) is Abelianised to 
U(1)$^3$ but it is still a self-interacting gauge theory. In \cite{g} the model 
(including a generalisation to Lorentzian signature and a cosmological constant) was 
solved in the continuum using a representation of the canonical commutation and adjointness
relations of Narnhofer-Thirring type \cite{h}. That is, the exponentiated constraints of
the theory could be defined as densely defined, in fact unitary, operators. In \cite{i}
the model was solved in Fock representations where the constraints of the theory are 
densely defined quadratic forms but no operators.    

In the present paper we follow the CQFT approach and consider the Hamiltonian 
renormalisation of the model. We do this both for the Narnhofer-Thirring and the 
Fock flow. In both cases we can compute the fixed point and find that it coincides with 
the known solutions \cite{g,i}.\\ 
\\
This work is organised as follows:\\

In section \ref{s2} we review the classical U(1)$^3$ model and the  
quantum solutions \cite{g,i}.     

In section \ref{s3} we perform the Hamiltonian renormalisation in the Narnhofer-Thirring
representation of \cite{g}. 

In section \ref{s4} we perform the Hamiltonian renormalisation in the Fock representation
representation of \cite{i}.

In section \ref{s7} we summarise and conclude.

In appendix \ref{sa} we have collected the renormalisation tools from 
\cite{l} which is related to wavelet theory \cite{m}.

\section{The U(1)$^3$ model}
\label{s2}

In the first subsection we present the bare bones of the classical Hamiltonian formulation
of the U(1)$^3$ model (see \cite{n} for a corresponding Lagrangian and Dirac constraint 
analysis) and we define what we mean by a successful quantisation.
In the second we review the solution of this model in the Narnhofer-Thirring 
representation following \cite{g}. In the third we review the solution of the model 
in the fock representation of \cite{i}.

\subsection{Classical Hamiltonian formulation and quantisation objective}
\label{s2.1}

The real phase space is coordinatised by a conjugate pair of fields $(A_a^j,E^a_j)$ on the 
spacetime manifold $\mathbb{R}\times \sigma$ where $\sigma$ is 3-manifold. For the purpose 
of this paper it will be sufficient to take $\sigma$ compact without boundary thus 
avoiding the boundary term analysis of \cite{o}. The spatial tensor indices take range 
$a,b,c .. =1,2,3$, the  u(1)$^3$ Lie algebra indices take range 
$j,k,l, .. =1,2,3$. Accordingly the Poisson brackets of the time zero fields or initial 
data are 
(we take Newton's constant and Planck's constant to be unity)
\be \label{2.1}
\{E^a_j(x),\;A_b^k(y)\}=\delta^a_b\;\delta^k_j\;\delta(x,y)
\ee
The phase space is subject to three types of constraints 
\be \label{2.3}
C_j=\partial_a E^a_j,\;
D_a=E^b\;(\partial_a A^j_b)-\partial_b (A_a^j E^b_j),\;
H=\epsilon_{jkl} [\partial_{[a} A_{b]}]\; E^a_k\; E^b_l\; |\det(E)|^{[w-2]/2}
\ee
known as the Gauss, spatial diffeomorphism and Hamiltonian constraint respectively.      
We consider $H$ in a chosen density weight $w$. 
Their smeared versions $C[r]=\int_\sigma\; d^3x\; r^j\; C_j$, 
$D[u]=\int_\sigma\; d^3x\; u^a\; D_a$ and 
$H[N]=\int_\sigma\; d^3x\; N\; H$ satisfy the hypersurface deformation algebroid 
relations
\ba \label{2.4}
&& \{C[r],C[s]\}=0,\; \{D[u],C[r]\}=-C[u[r]],\;\{C[r],H[N]\}=0,
\\
&& \{D[u],D[v]\}=-D[[u,v]],\; \{D[u],H[N]\}=-H[u[N]],
\{H[N],H[N']\}=-D[Q(N\; dN'-N'\; dN)]
\nonumber
\ea
where $Q^{ab}=E^a_j E^b_k \delta^{jk} |\det(E)|^{2(w-2)}$ depends on the density 
weight and $u[s], [u,v], u[N]$ denote the Lie derivatives of $s,v,N$ with respect
to the vector field $u$ and $s,v,N$ are considered respectively as scalar, vector field and 
scalar of density weight $-(w-1)$. Note that the classical Dirac analysis naturally 
selects $w=1$ which means that the classical phase space is constrained by $\det(E)\not=0$
(non-degeneracy \cite{q}).
By a quantum solution we mean 1. a quantisation of the canonical Poisson brackets
(\ref{2.1}) and the reality conditions, stating that $A,E$ are real valued, 
as well as 2. some version of (\ref{2.4}). A convenient way to state this precisely 
is to construct first a Weyl algebra $\mathfrak{A}$ generated by the Weyl elements 
\be \label{2.5}
W[F]=e^{-i<F,A>},\; W[G]=e^{-i<G,E>};\;\;
A[F]:=<F,A>=\int_\sigma\; d^3x\; F^a_j\; A_a^j,\;
E[G]:=<G,E>=\int_\sigma\; d^3x\; G_a^j\; E^a_j,\; 
\ee
where $(F,G)$ are taken from a space ${\cal S}$ of real valued smearing functions sufficiently general 
in order that the $W[F], \; W[G]$ separate the points of the classical phase space.
The Weyl relations read
\begin{align} \label{2.6}
&W[G]\; W[F]\; W[-G]=e^{-i<G,F>_L}\;W[F],\; 
W[F]\;W[F']=W[F+F'],\;W[G]\;W[G']=W[G+G'],
\nonumber\\
& W[0]=1_{\mathfrak{A}},\;
W[F]^\ast=W[-F],\; W[G]^\ast=W[-G] 
\end{align}
It is important to note that different choices of $\cal S$ generate different 
$\mathfrak{A}$. Namely, the requirement that the $W[F], \; W[G]$ are represented by well
defined (in fact unitary) operators on a Hilbert space implies that 
the unsmeared fields take values in the dual space  ${\cal S}^\ast$ of distributions. 
In case that  $\cal S$ is equipped with a topology, we take ${\cal S}^\ast$ as 
the topological dual (continuous linear functionals) otherwise the algebraic dual 
(just linear functionals). 

Then a cyclic representation $(\rho, {\cal H}, \Omega)$ of $\mathfrak{A}$ is is in 1-1 correspondence
with a state $\omega$ 
(positive, linear, normalised functional) on $\mathfrak{A}$ via the GNS construction 
\cite{p} up to unitary equivalence. Here $\rho$ is a representation of $\mathfrak{A}$
by operators on the Hilbert space $\cal H$ and $\Omega$ is a vector such that 
${\cal D}=\rho(\mathfrak{A})\;\Omega$ is dense. The correspondence is via
$\omega(a)=<\Omega,\; \rho(a)\; \Omega>_{{\cal H}}$ for all $a\in \mathfrak{A}$ 
($a$ is any finite linear combination of Weyl elements with complex valued coefficients). 
Thus a state solves the first task 1. of the quantisation problem. The second task 2.
is to also represent the $C[r],\; D[u],\; H[N]$ or perhaps their exponentials 
$e^{i\;C[r]},\; e^{i\;D[u]},\; e^{i\;H[N]}$ by operators on $\cal H$ such that 
(\ref{2.4}) or an exponentiated version thereof is implemented by replacing Poisson brackets 
divided by the imaginary unit. The non-exponentiated version would be a quantum 
realisation of the algebroid while  the exponentiated version would be a quantum 
realisation of the corresponding groupoid.

\subsection{Groupoid solution}
\label{s2.2}

For that solution \cite{g} we consider the following choices:\\
1. ${\cal S}\subset [C^\infty(\sigma)]^9 \times [C^\infty(\sigma)]^9$ i.e. both smearing 
fields of the Weyl algebra take values in the set of smooth functions (which could be 
equipped with some topology but that will not be important for what follows).\\
2. We pick general $w$. The choice $w=2$ has the advantage that all constraints have minimal
polynomial degree. \\
3. We pick the Narnhofer-Thirring type of state on the corresponding $\mathfrak{A}$
\be \label{2.7}
\omega(W[F]\;W[G])=\delta_{F,0}
\ee
This choice means that the state is regular with respect to $G$ but not with respect to 
$F$, i.e. the operator $\rho(E[G])$ exists and in fact annihilates the GNS vacuum $\Omega$
but the operator $\rho(A[F])$ does not exist, only its exponential $\rho(W[F])$ does. 
Moreover the corresponding GNS Hilbert space is not separable and has an ONB consisting of 
the vectors $\rho(W[F])\Omega$.\\ 
4. We choose to represent the exponentials of the constraints as operators.\\
\\
The latter step is non-trivial and requires a regularisation of the constraints 
$D[u], H[N]$ which are written in terms of $A_a^j(x)$ and not in terms of the $W[F]$.
Since $A_a^j(x)$ does not exist in the chosen representation, one must write it 
as a limit of an expression involving the $W[F]$ and at the end take the regulator  
away. The details can be found in \cite{g}. For the purpose of the present paper 
it will be sufficient to proceed formally and check that the end result is well
defined in the chosen representation and displays a suitable representation of the groupoid.

We pick the following factor ordering
\ba \label{2.8}
C[r] &:=& \int\; d^3x\; C_a^j(r)\; E^a_j:=\int\; d^3x\;[-r^j_{,a}]\; E^a_j
\nonumber\\
D[u] &:=& \int\; d^3x\; A_a^j\; D^a_j(u,E),\; D^a_j(u,E):=u^a_{,b} E^b_j-(u^b E^a_j)_{,b},\;
\nonumber\\ 
H[N] &:=& \int\; d^3x\; A_a^j\; H^a_j(N,E),\; H^a_j(N,E):=\epsilon_{jkl}\;(N\;E^{[a}_k E^{b]}_l\;|\det(E)|^{[w-2]/2})_{,b}
\ea
The relation $E[G] \; W[F]=W[F] [E[F]+<G,F> 1_{\mathfrak{A}}]$ which follows from the 
Weyl relations implies using $\rho(E[G])\Omega=0$ that the $\rho(W[F])\Omega$ are eigenstates 
of $\rho(E^a_j(x))$ with eigenvalue $F^a_j(x)$. Accordingly we define 
\be \label{2.9}
\rho(D[u]) \; \rho(W[F])\Omega=[\int\; d^3x\; \rho(A_a^j)\; D^a_j(u,F)]\; \rho(W[F])\Omega
\ee
Since formally $\rho(W[F]))=\exp(-i<F,\rho(A)>)$ making use of the fact that $\rho$ is a
$\ast-$homomorphism  $\rho(a+b)=\rho(a)+\rho(b),\;\rho(a\;b)=\rho(a)\;\rho(b),\;
\rho(a^\ast)=[\rho(a)]^\dagger$ for all $a,b\in \mathfrak{A}$ allows us to rewrite (\ref{2.9}})
as 
\be \label{2.10}
\rho(D[u]) \; \rho(W[F])\Omega=i[\int\; d^3x\; D^a_j(u,F)]\;\frac{\delta}{\delta F^a_j} \rho(W[F])\Omega  
=:i\; <D(u,F), \frac{\delta}{\delta_F}>\; \rho(W[F])\Omega
\ee
where $\delta/\delta F$ is the functional derivative. Note that the ordering of $D^a_j$ and 
$\delta/\delta F^a_j$ displayed is mandatory. 
We emphasise that in this step it was important that $F$ is sufficiently general to separate the 
points. This will have an important consequence for the renormalisation procedure in later sections.
Proceeding formally we find for the exponentiated constraint
\ba \label{2.11}
&& \rho(e^{-i D(u)})\; \rho(W[F])\Omega=e^{-i \rho(D(u))}\; \rho(W[F])\Omega
=e^{-<D(u,F), \frac{\delta}{\delta_F}>}\; e^{-i<F,\rho(A)>}\Omega
\nonumber\\
&=&  e^{<D(u,F), \frac{\delta}{\delta_F}>}\; e^{-i<F,\rho(A)>}\; e^{<D(u,F), \frac{\delta}{\delta_F}>}\Omega
= \exp(-i<e^{<D(u,F), \frac{\delta}{\delta_F}>}\; F\; \; e^{-<D(u,F), \frac{\delta}{\delta_F}>},\;\rho(A)>\;\Omega
\nonumber\\
&=& W[e^{<D(u,F), \frac{\delta}{\delta_F}>}\; F\; \; e^{-<D(u,F), \frac{\delta}{\delta_F}>}]\;\Omega
\ea
where in the second line we used that the vacuum is independent of $F$, that is, $\delta/\delta F\; \Omega=0$.
While the intermediate steps require a regularisation procedure, the end result of (\ref{2.11}) 
is well defined. In fact, it has a simple geometrical interpretation: Let $K^a_j(A,E):=E^a_j$ the 
momentum coordinate function on phase space. Then  
\be \label{2.12}
e^{<D(u,F), \frac{\delta}{\delta_F}>}\; F\; \; e^{-<D(u,F), \frac{\delta}{\delta_F}>}
=[e^{-X_u}\cdot K](0,F)
\ee
where $X_u$ is the Hamiltonian vector field of $D[u]$. 

For the constraint $H[N]$ we can proceed in exactly the same fashion because all the above steps 
just relied on the constraint being linear in $A$. Hence in terms of the Hamiltonian vector 
field $X_N$ of $H[N]$ we find 
\be \label{2.13}
\rho(e^{-i\; H[N]})\;\rho(W[F])\Omega=W[(e^{-X_N}\cdot K)(0,F)]\;\Omega
\ee
The exponentiated Gauss constraint is in fact diagonal
\be \label{2.14}
\rho(e^{-i\; C[r]})\;\rho(W[F])\Omega=e^{i\; <F,dr>}\;W[F]\;\Omega
\ee
These operators are densely defined on $\cal D$ and in fact unitary as long as 
$X_u, X_N$ are well defined which for $w<2$ imposes that we require $\det(F)\not=0$. 

It remains to verify (\ref{2.4}). The irregularity of the representation of $W[F]$  with respect 
to $F$ is transported into $D[u], H[N]$, i.e. $t\mapsto \rho(e^{i\;t\;D[u]}),\; \rho(e^{i\;t\;H[N]})$
are 1-parameter unitary groups but they are not strongly continuous and we cannot verify 
(\ref{2.4}) in its non-exponentiated form. As a substitute we have
\ba \label{2.15}
&& \rho(e^{i\;\;D[u]})\; \rho(e^{i H[N]}) \;\rho(e^{-i\;\;D[u]})\; W[F]\Omega
=W[(e^{-X_u}\cdot e^{X_N}\cdot e^{X_u}\cdot K)(0,F)]\;\Omega
\nonumber\\
&=& W[\exp(e^{-X_u}\cdot X_N\cdot e^{X_u})\cdot K)(0,F)]\;\Omega  
=W[\exp(X_{e^{L_u}\cdot N})\cdot K)(0,F)]\;\Omega
\nonumber\\   
&=& \rho(e^{i H[e^{L_u}\cdot N]}) \; W[F]\Omega,
\ea
where $L_u$ denotes the Lie derivative and we made use of the homomorphism property 
$[X_A,X_B]=X_{\{A,B\}}$ of Hamiltonian vector fields of phase space functions $A,B$.
Likewise 
\be \label{2.16}
\rho(e^{i\;\;D[u]})\; \rho(e^{i D[v]}) \;\rho(e^{-i\;\;D[u]})\; W[F]\Omega
=\rho(e^{i D[e^{L_u}\cdot v]}) \; W[F]\Omega.
\ee
Finally
\be \label{2.17}
\rho(e^{i\;\;H[M]})\; \rho(e^{i H[N]}) \;\rho(e^{-i\;\;H[M]})\; W[F]\Omega
=W[\exp(e^{-X_M}\cdot X_N\cdot e^{X_M})\cdot K)(0,F)]\Omega,
\ee
which qualifies as a quantisation of $e^{-X_M}\cdot H[N] \; e^{-X_M}$ since 
this expression is also linear in $A$.

\subsection{Algebroid solution}
\label{s2.3}

For that solution \cite{i} we consider the following choices:\\
1. ${\cal S}\subset [C^\infty(\sigma)]^9 \times [C^\infty(\sigma)]^9$ as in the previous subsection.\\
2. We pick $w=2$ so that all constraints have minimal
polynomial degree (namely three). \\
3. We pick the Fock state $\omega$ such that the cyclic vector $\Omega$ is annihilated by 
the annihilator $a_a^j=2^{-1/2}[A_a^j-i\; \delta_{ab} \delta^{jk} E^b_k]$, i.e.
\be \label{2.18}
\omega(W[F]\; W[G])
=e^{-\frac{1}{4}[<F,F>+<G,G>]+\frac{i}{2} <F,G>}
\ee
where $<F,F>=\int\; d^3x \;\delta_{ab}\delta^{jk} F^a_j F^b_k$,
$<G,G>=\int\; d^3x \;\delta^{ab}\delta_{jk} G_a^j G_b^k$ and 
$<F,G>=\int\; d^3x \;F^a_j\; G_a^j$. In contrast to the previous section,
this representation is constructed using the flat spatial background metric $\delta_{ab}$.
On the other hand $\omega$ is regular with respect to both $F,G$ so that $A,E$ exist 
as operator valued distributions and the Fock Hilbert space is separable spanned 
by the Fock vectors $\Omega\, <F_1,a>^\ast\;..\;<F_N,a>^\ast \Omega,\; N=1,2,..$.\\
4. We choose to represent the (\ref{2.4}) as quadratic forms.\\
\\
The latter step is not difficult: Being polynomials in $A,E$ we write 
$A=2^{-1/2}[a+a^\ast],\; E=i\;2^{-1/2}[a-a^\ast]$ and normal order, hence 
\be \label{2.19}
\rho(C[r])=:C[r,\rho(a),\rho(a)^\dagger]:,\;
\rho(D[u])=:D[u,\rho(a),\rho(a)^\dagger]:,\;
\rho(H[N])=:H[N,\rho(a),\rho(a)^\dagger]:,\;
\ee
However, what is difficult is to verify (\ref{2.4}) because the objects 
(\ref{2.19}) are merely quadratic forms but not operators. E.g. 
$||\rho(D[u])\Omega||=\infty$ while matrix elements of $\rho(D[u])$ between 
Fock vector states are well defined. To deal with this problem we follow
\cite{i} and introduce a real valued, smooth orthonormal basis $b_I$ of the 
one particle Hilbert space $\mathfrak{h}:=L_2(\sigma, d^3x)$ where $I\in {\cal I}$ is a countable
index set of modes. For the important case $\sigma=T^3$ considered in the section 
of renormalisation we may pick ${\cal I}=\mathbb{Z}^3$ and the functions 
$b_{\vec{n}}(x)=\prod_{a=1}^3\; b_{n_a}(x^a)$ where 
$b_0=1,\; b_n(x)=2^{1/2}\cos(2\pi\;n\;x)\; (n>0),\;b_n(x)=-2^{1/2}\sin(2\pi\;n\;x)\; (n<0)$.
We consider a function $|.|: {\cal I}\mapsto \mathbb{N}$ which has the property that the sets 
${\cal I}_M=\{I\in {\cal I}; |I|\le M\}$ are nested, that is, ${\cal I}_M\subset {\cal I}_{M'}$ 
when $M<M'$. For $T^3$ we may pick e.g. $|\vec{n}|={\sf max}(\{|n_a|;\;a=1,2,3\})$. 
We now use the resolution of identity $a_a^j(x)=\sum_{I\in {\cal I}}\; b_I(x)\; a_a^j(I)$
where $a_a^j(I):=<b_I,a_a^j>_{\mathfrak{h}}$
and substitute this into (\ref{2.19}). In this way $\rho(C[r]),\;\rho(D[u]),\;  \rho(H[N])$ 
become respectively single, double and triple infinite sums over $\cal I$ with respect 
to the $a_a^j(I)$ and $a_a^j(I)^\dagger$. By 
$\rho(C[r])_{M_1},\;  \rho(D[u])_{M_1,M_2},\;  \rho(H[N])_{M_1,M_2,M_3}$ we mean 
the truncation of the first, second and third sum respectively to the sets 
${\cal S}_{M_1},\; {\cal S}_{M_2},\; {\cal S}_{M_3}$ respectively. These truncations 
are now well defined operators on the Fock space and their commutators can be computed. 

We say that those commutators have a limit as quadratic forms iff there is a limiting pattern
in which one can take the respective $M_1,M_2,M_3$ to infinity in the weak operator topology.
E.g. we take any Fock states $\psi,\psi'\in {\cal H}$ and ask whether it is 
possible to send $M_1, M_2, M_1', M_2'$ in 
\be \label{2.20}
<\psi,   [\rho(D[u])_{M_1,M_2},\rho(D[v])_{M_1',M_2'}] \psi'>_{{\cal H}} 
\ee 
to infinity, defining the matrix elements of a well defined quadratic form. The sequence or 
pattern in which we take $M_1,M_2,M_1', M_2'$ to infinity is part of the definition of that quadratic form.
Since on Fock space only normal ordered expressions can be well defined, in (\ref{2.20})
one has to restore normal order which does not produce divergences at finite $M_1,..,M_2'$
but there are normal reordering contributions in both products 
$\rho(D[u])_{M_1,M_2}\;\rho(D[v])_{M_1',M_2'}$ and 
$\rho(D[v])_{M_1',M_2'},\rho(D[u])_{M_1,M_2}$ which diverge individually. Now the task is to 
show that these divergences can be made to cancel when we subtract these two products 
in the commutator by a judicious choice of limiting pattern. In \cite{i} it was shown that 
indeed such a limiting pattern can be found for each of the six commutators between the 
(\ref{2.19}) such that (\ref{2.4}) is realised without anomalies. In particular
\be \label{2.21}
[\rho(H[M]),\rho(H(N))]=i\;:\{H[M],H[N]\}(\rho(A),\rho(E)):    
\ee 
in the sense of quadratic forms. The r.h.s. has to be understood in the following way:
take the classical Poisson bracket $\{H[M],H[N]\}(A,E)$, substitute 
$A,E$ by $2^{-1/2}[\rho(a)+\rho(a)^\dagger],\; i\;2^{-1/2}[\rho(a)-\rho(a)^\dagger]$ 
and normal order.

\section{The groupoid flow}
\label{s3}

We consider Hamiltonian renormalisation of the continuum theory presented in section 
\ref{s2.2} using the renormalisation tools listed in appendix \ref{sa}, thereby specialising 
to $\sigma=T^3$.
The section is organised as follows. In the first subsection we use projector maps 
$P_M$ -as defined in \cref{sa}-, to project from $L$ to the $L_M$ subspaces and then compute the renormalisation flow. 
Here we work with general density weight $w$ in the constraint algebra. We encounter a subtlety that 
draws its origin from the discontinuity of the Narnhofer-Thirring representation. We use a toy model in the 
second subsection to explain the mechanism at work in non-technical terms.   
In the third subsection we discretise the fields, work in the $l_M$ spaces of square summable sequences and 
compute the renormalisation flow for the Hamiltonian constraint with polynomial weight $w=2$. The reason for performing 
both of these essentially equivalent renormalisations is that the first flow can be considered as a flow of smeared 
continuum fields outside of a lattice context while the second is more in the tradition of the ``real space'' or lattice block spin 
flow of discretised fields. The real space perspective suggests a different, apparently more local  
starting point for the flow equations and converges less rapidly to the correct fixed point which displays essential
non-localities. Finally, in the fourth subsection we summarise and compare our findings and relate  
these to the perspectives of the actual SU(2) theory of Euclidan signature quantum gravity.

\subsection{Renormalisation flow for projected constraints}
\label{s3.1}

Using the tools developed in \cref{sa} we start by considering the projected fields at resolution scale $M$
\be \label{3.1}
A_{M,a}^j(x)=\int_{T^3}\; d^3x\; P_M(x,y)\; A_a^j(y),\;\;\; 
E^{a}_{M,j}(x)=\int_{T^3}\; d^3x\; P_M(x,y)\; E^a_j(y),\;
\ee
and similarly $G_{M, a}^j(x),\; F^{a}_{M,j}(x)$. A clarification on the notation is in order: 
whenever a comma appears between \( M \) and a space or internal index, it does \emph{not} indicate differentiation. 
In contrast, a comma placed at the far right of an expression will, as usual, denote a partial derivative. 
The index $M$ will remind us that at which resolution we are working.\\  
\\
The non-vanishing Poisson brackets 
are 
\be \label{3.2}
\{E^{a}_{M,j}(x), A_{M, b}^k(y)\}=\delta^a_b\;\delta^k_j\; P_M(x,y),
\ee
and the Weyl elements are given by 
\be \label{3.3}
W_M[F_M]=e^{-i<F_M,A_M>_{L_M}},\;W_M[G_M]=e^{-i<G_M,E_M>_{L_M}}.
\ee
Using the fact that $P_M^2=P_M$ is a projection we also have 
$W_M[F_M]=W[F_M],\; W_M[G_M]=W[G_M]$ where $W$ are the continuum Weyl elements 
defined in section \ref{s2.2}. The Weyl algebra $\mathfrak{A}_M$ is generated 
by (\ref{3.3}) using the Weyl relations
\begin{align} 
\label{3.4}
& W_M[G_M]\; W_M[F_M]\; W_M[-G_M]=e^{-i<G_M,F_M>_{L_M}}\;W_M[F_M],\nonumber \\ & 
W_M[F_M]\; W_M[F_M']=W_M[F_M+F'_M],\;\;
W_M[G_M]\; W_M[G_M']=W_M[G_M+G'_M],\;
\nonumber\\
& W_M[0]=1_{\mathfrak{A}_M},\; W_M[F_M]^\ast=W_M[-F_M],\; W_M[G_M]^\ast=W_M[-G_M].
\end{align}
To initialise the renormalisation flow we pick for each $M$ the state 
\be \label{3.5}
\omega^{(0)}_M(W_M[F_M]\; W_M[G_M]):=\delta_{F_M,0}
\ee
which defines a cyclic representation $\rho^{(0)}_M$ of $\mathfrak{A}_M$ on ${\cal H}^{(0)}_M$ with 
cyclic vector $\Omega^{(0)}_M$. Note that again $\rho^{(0)}_M(E_M[G_M])$ is diagonal on 
$\rho^{(0)}_M(W_M[F_M])\Omega^{(0)}_M$ with eigenvalue $<F_M,G_M>_{L_M}$. 

Following the general recipe of appendix \ref{sa} the classical constraints at cut-off resolution $M$ are given by 
\ba \label{3.6}
C_M[r] &=& -\int\;d^3x\; r^j_{,a} \;E^{a}_{M,j}
\nonumber\\
D_M[u] &=& \int\; d^3x\; A_{M,a}^j\; D^a_j(u, E_M),\;\;\; D^a_j(u,E_M):=u^a_{\;,b}\; E^{b}_{M,j}-(u^b E^{a}_{M,j})_{,b},\;
\nonumber\\ 
H_M[N] &:=& \int\; d^3x\; A_{M\;a}^j\; H^a_j(N,E_M),\;\;\; H^a_j(N,E_M):=\epsilon_{jkl}\;(N\; E^{[a}_{M,k} E^{b]}_{M,l}
|\det(E_M)|^{(w-2)/2})_{,b}
\ea
Using the results discussed in \cref{s2.2} we represent quantisations of the exponentials of (\ref{2.6}) on the dense subspace 
${\cal D}^{(0)}_M=\rho^{(0)}_M(\mathfrak{A}_M)\Omega^{(0)}_M$ by 
\ba \label{3.7}
&& \rho^{(0)}_M(e^{i\;C_M[r]},c^{(0)}_M) \rho^{(0)}_M(W_M[F_M])\Omega^{(0)}_M:=e^{-i<F_M,dr>} \; \rho^{(0)}_M(W_M[F_M])\Omega^{(0)}_M
\nonumber\\
&&\rho^{(0)}_M(e^{i\;D_M[u]},c^{(0)}_M)\; \rho^{(0)}_M(W_M[F_M])\Omega^{(0)}_M:=W_M[(e^{X^M_u} \cdot K)(0,F_M)]\Omega^{(0)}_M
\nonumber\\
&& \rho^{(0)}_M(e^{i\;H_M[N]},c^{(0)}_M)\; \rho^{(0)}_M(W_M[F_M])\Omega^{(0)}_M:=W_M[(e^{X^M_N} \cdot K)(0,F_M)]\Omega^{(0)}_M
\ea
Here, as in section \ref{s2.2}, $K$ is the momentum coordinate function on the continuum phase space 
given by $K^a_j(A,E)=E^a_j$.  $X^M_u, X^M_N$ are the Hamiltonian vector fields of $D_M[u],\; H_M[N]$ 
respectively, considered as functions on the continuum phase space and $c^{(0)}_M$ are functions that 
parametrise the discretisation choices at resolution $M$. Explicitly (for $w=2$)
\ba \label{3.8}
&& D_M[u](A,E)=
\int\;d^3x\; A_a^j(x)\int\; d^3y\; P_M(x,y)\;\int\; d^3z\;  [u^a_{,b}(y) P_M(y,z)-\delta^a_b\;(u^c(y) P_M(y,z))_{,y^c}] \; E^b_j(z) 
\nonumber\\
&& H_M[N](A,E)=
\int\;d^3x\; A_a^j(x)\int\; d^3y\; P_M(x,y)\;\int\; d^3z_1\;\int\; d^3z_2\; \epsilon_{jkl} \;  
\times\nonumber\\
&& (N(y) \delta^{[a}_c \delta^{b]}_d\; P_M(y,z_1)\; P_M(y,z_2))_{,y^b}\; E^c_k(z_1)\; E^d_l(z_2)
\ea
These quantisations are motivated by following the exact same formal steps as in section 
\ref{s2.2}. In more detail, 
for instance
\ba \label{3.8a}
&& \rho^{(0)}_M(D_M[u])\; \rho^{(0)}_M(W_M[F_M])\Omega^{(0)}_M
=[\int\; d^3x\;\rho^{(0)}_M(A_{M,a}^j(x))\; D^a_j(u, F_M)(x)]\;\rho^{(0)}_M(W_M[F_M])\Omega^{(0)}_M
\nonumber\\
&=& \{i[\int\; d^3x\; D^a_j(u, P_M\cdot F)(x)\; (P_M\cdot \frac{\delta}{\delta F^a_j})(x)]\;
[\rho^{(0)}_M(W_M[F])]\Omega^{(0)}_M\}_{F=F_M}
\ea
where the formal extension $W_M[F]=e^{-i<F,A_M>}$ was defined and 
we made use of $P_M^2 =P_M$. From here on the computation is identical to that of section (\ref{s2.2}).
Instead of working with this formal extension  
we can make use of the bijection between $L_M, l_M$ and introduce, given 
$F_M$, the discrete function $f_M=I_M^\dagger\; F_M\;\; \Leftrightarrow\;\; F_M=I_M\; f_M$. Then 
for any functionally differentiable functional $K[F]$ with restriction $K[F_M]$ we have the identity
\be \label{3.8b}
\{[(P_M\cdot \frac{\delta}{\delta F^a_j})(x) K][F]\}_{F=F_M} 
=\sum_{m\in \mathbb{N}_M^3}\; \chi^M_m(x)\; \frac{\partial}{\partial f^a_{M,j}(m)}\; K[F_M]
=:\frac{\delta}{\delta F^a_{M,j}(x)}\; K[F_M] 
\ee
where $M^{-3}\sum_m \chi^M(x)\chi^M_m(y)=P_M(x,y)$ and the chain rule 
$\delta K[F_M]=\int\; d^3y\; [\frac{\delta H[F]}{\delta F^a_j(y)}]_{F=F_M}\; (\delta F^a_{M,j})(y)$ 
with $\delta F^a_{M,j}(y)=M^{-3}\sum_m \;\chi_m^M(y)\; \delta f^a_{M,j}(m)$
was used. In particular $\frac{\delta F^a_{M,j}(x)}{\delta F^b_{M,k}(y)}=P_M(x,y)\delta^a_b\delta^k_j$. 

We now compute the flow of the initial family (\ref{3.5}) and (\ref{3.7}). We have by definition
\be \label{3.9}
\omega^{(1)}_M(W_M[F_M]\; W_M[G_M]):=
\omega^{(0)}_{3M}(W_{3M}[F_M]\; W_{3M}[G_M])
=\delta_{F_M,0}
\ee
where $F_M,G_M$ are considered as elements of $L_{3M}$ since $L_M\subset L_{3M}$ is a subspace.
Accordingly 
\be \label{3.10}
\omega^{(1)}_M=\omega^{(0)}_M=\omega^{(n)}_M=\omega^\ast_M
\ee
is already fixed pointed and indeed 
\be \label{3.11}
\omega^\ast_M(W_M[F_M]\; W_M[G_M])=\omega(W[F_M]\; W[G_M])
\ee
where $\omega$ is the continuum state of section \ref{s2.2}. We drop the label $\ast$ 
from $\omega^\ast_M$ and correspondingly from the GNS data 
$(\rho_M, {\cal H}_M, \Omega_M)$ in what follows for better readability, $\omega_M$ is simply 
the restriction of $\omega$ to $\mathfrak{A}_M$.\\
\\
We now consider the constraints. Here we encounter a new effect which arises due to 
the extreme discontinuity of the Narnhofer-Thirring representations with respect to the 
labels $F,F_M$ which is not present in regular representations such as Fock representations
and which requires to adapt the definition of the flow equations as we will see. Consider e.g. the
flow equation for the spatial diffeomorphism constraint in the first iteration step, according
to its definition derived for regular representations. It  would read 
\ba \label{3.12}
&& <\rho_M(W_M[F'_M])\Omega_M,\; \rho^{(1)}_M(e^{i D_M[u]},c^{(1)}_M) \; \rho_M(W_M[F_M])\Omega_M>_{{\cal H}_M}
\nonumber\\
&=&
<\rho_{3M}(W_{3M}[F'_M])\Omega_{3M},\; \rho^{(0)}_{3M}(e^{i D_{3M}[u]},c^{(0)}_{3M}) \; \rho_{3M}(W_{3M}[F_M])\Omega_{3M}>_{{\cal H}_{3M}}
\ea
where again the coefficients $c^{(k)}_M$ denote the quantisation choices to be made at resolution $M$ at the 
$k-$th iteration. However we see that the right hand side trivially vanishes because the smearing 
function $[e^{X^{3M}_u}\cdot K](F_M)$ no longer lies in in $L_M$. Therefore the flow would return 
zero for all exponentiated constraints as the fixed point which is clearly non-sensical.\\
\\
In what follows we first unveil the mechanism behind this observation and adapt the 
flow equations. Then we consider that adapted flow for the constraints truncated in terms 
of projections $P_M$ and after that in terms of discretisations $I_M^\dagger$. 
These two options are strictly equivalent. The projection formalism is more in the spirit 
of constructive QFT in the sense that one smears the operator valued distributions with 
respect to smooth test functions, the discretisation formalism emphasises the spatial 
quasi-locality of the constraints and makes the renormalisation procedure resemble the classical
``real space'' block spin transformations. The discretisation formalism suggests more general 
``naive'' discretisations than those obtained via projection and indeed makes the flow 
equations less trivial in that case. Therefore for sake of clarity we discuss the flow 
in terms of projections for general density weight $w$ while for the flow in terms of 
discretisations we discuss it for $(w-2)/2$ an even non-negative integer so that the 
Hamiltonian constraint is a polynomial.

\subsubsection{Toy model}
\label{s3.1.0}

To understand the origin of this effect in non-technical terms and why the prescription 
(\ref{3.12}) needs to be adapted accordingly we consider the following simple toy model:
Consider an only 2-dimensional phase space with canonical pairs $(A_I,E^I),\; I=1,2$ and 
an Narnhofer-Thirring representation of the corresponding Weyl algebra, that is 
\be \label{3.a}
\omega(W[F]\; W[G])=\delta_{F,0},\; W[F]=e^{-i\; F^I\; A_I},\; W[F]=e^{-i\; G_I\; E^I},\; 
\ee
as well as a classical constraint ($f_I$ are general smooth functions)
\be \label{3.b}
C_u=u^{IJ}\; A_I\;f_J(E^1,E^2)
\ee
The representation is smooth with respect to $G$ but totally discontinuous with respect to
$F$. If $\Omega$ is the GNS vacuum then in the GNS representation $\rho$ the operators
$\rho(E^I)$ are diagonal with eigenstates $\rho(W[F])\Omega$ and eigenvalue $F^I$. Dropping 
the $\rho$ function for notational simplicity the naive action of the quantisation of (\ref{3.b}) 
on the 
orthonormal basis $W[F]\Omega$ is therefore given by 
\be \label{3.c}
-i\tilde{C}_u\; W[F]\Omega=-i\;u^{IJ}\; A_I\;f_J(F^1,F^2)\; W[F]\Omega
\ee  
However, due to the irregularity, $A_I$ is not an operator, only Weyl elements are. 
Hence (\ref{3.c}) is in fact ill-defined and we must regularise it. Given $\epsilon>0$ 
we define the regulated constraint by
\be \label{3.d}
-i\;C^\epsilon_u\; W[F]\Omega=u^{IJ}\; f_J(F^1,F^2)\; (\partial^\epsilon_I\; W)[F]\Omega
\ee     
where for a general functional $K[F]$ we define using the vector $\delta_I$ with 
components $(\delta_I)^J=\delta^J_I$
\be \label{3.e}
(\partial^\epsilon_I \; K)[F]:=\frac{1}{\epsilon}(K[F+\epsilon \delta_I]-K[F+\epsilon^2 \delta_I])
\ee
If we would use the standard topology of $\mathbb{R}^2$ and for smooth $H$ the limit 
$\epsilon\to 0$ of (\ref{3.e}) would simply produce the partial derivative 
$\partial_I$ and in that sense (\ref{3.d}) would return (\ref{3.c}) in the limit. However,
that limit cannot be taken because in the weak operator topology as the vectors (\ref{3.d})
for different $\epsilon$ are mutually orthogonal. On the other hand, (\ref{3.d}) is a 
linear combination of Weyl elements and therefore well defined. 

To exponentiate (\ref{3.d}) we introduce the operator acting on the $F$ labels
\be \label{3.f}
(X^\epsilon_u\; K)[F]:=u^{IJ}\; f_J(F^1,F^2)\; (\partial^\epsilon_I\; K)[F]
\ee
Then using formally the Baker-Campbell-Hausdorff formula
\be \label{3.g}
e^{-i\;C^\epsilon_u}\; W[F]\Omega=W[(e^{X^\epsilon_u}\; K)[F]]\;\Omega 
\ee
where the coordinate functionals $K^I[F]:=F^I$ have been introduced and we exploited that
$\Omega$ is independent of $F$. The point is now 
while still the vectors (\ref{3.g}) are mutually orthogonal for different $\epsilon$ we 
can take the limit $\epsilon\to 0$ of $(e^{X^\epsilon_u}\; K)[F]$ in the topology 
of smooth functions and obtain $(e^{X_u}\; K)[F]$ where 
\be \label{3.h}
(X_u\; K)[F]:=u^{IJ}\; f_J(F^1,F^2)\; (\partial_I\; K)[F]
\ee
This is then the motivation to {\it define}
\be \label{3.i}
e^{-i\;C_u}\; W[F]\Omega:=W[(e^{X_u}\; K)[F]]\;\Omega 
\ee

We now consider the subspace ${\cal H}_1$ obtained as the closed linear span of the 
$W[F]\;\Omega$ with $F^2=0$. Thinking of the indices $I$ as truncation labels the classical 
truncated or projected constraint would read 
\be \label{3.j}
C_{1,u}=u^{1J}\; A_1\; f_J(E^1,0)
\ee
Its quantisation follows exactly the same steps as before except that everything is reduced 
to only one canonical pair
\be \label{3.k}
e^{-i\;C_{1,u}}\; W_1[F^1]\Omega_1=W[(e^{X_{1,u}}\; K_1)[F^1]]\;\Omega_1 
\ee 
where $W_1[F^1]=W[F^1,0],\; K_1[F^1]=F^1=K^1[F^1,F^2],\; \Omega_1=\Omega$. On the other hand, we 
can consider (\ref{3.i}) which was obtained for generic  $F^1, F^2$ and take the 
formal limit $F^2\to 0$ (this limit is not in the weak topology but in the discrete one, 
that is, it is simply the restriction)
\be \label{3.l}
e^{-i\;C_u}\; W[F^1,0]\;\Omega:=W[(e^{X_u}\; K)[F^1,0]]\;\Omega 
\ee
We have 
\be \label{3.m}
(X_u\; K^I)[F]=u^{IJ}\; f_J[F],\; (X_u^2\; K^I)(F)=u^{KL}\; f_L[F]\;\partial_K\;(u^{IJ} f_J[F])
\ee
etc. Evaluating (\ref{3.m}) at $F^2=0$ returns a non-zero result for $I=2$ unless $u^{2J}=0$. It 
follows that matrix elements of (\ref{3.l}) between states $W[\tilde{F}^1,0]\Omega,\;
W[F^1,0]\Omega$ trivially vanish unless $u^{2I}=0$. Hence, defining a flow equation
\be \label{3.n}  
<W_1[\tilde{F}^1]\Omega_1,\; e^{-i C_{1,u}}\; W_1[F^1]\Omega_1>:=
<W[\tilde{F}^1,0]\Omega,\; e^{-i C_u}\; W[F^1,0]\Omega>
\ee
fails whenever $u^{2J}\not=0$. 

The reason for this effect is the lack of continuity of the representation. The derivation 
of (\ref{3.i}) assumed that we worked on the full Hilbert space in the sense that 
we could have used above arguments in the sense of matrix elements 
\be \label{3.o}
<W[\tilde{F}]\Omega,\; e^{-i\;C_u}\; W[F]\Omega>:=<W[\tilde{F}]\Omega,\; W[(e^{X_u}\; K)[F]]\;\Omega> 
=\delta_{\tilde{F}, (e^{X_u}\; K)[F]}
\ee
which yields a non-trivial result. However, due to weak discontinuity, we do not expect this 
to have a limit at $F^2=\tilde{F}^2=0$ which would qualify as the proper quantisation of the matrix 
elements  of the exponential of $-iC_u$ between states $W[F^1,0]\tilde{\Omega},\; W[F^1,0]\Omega$.
In other words, projecting the regulated constraint to ${\cal H}_1$ and exponentiation do not commute. 
In continuous representations the two processes would commute, because via Stone's theorem 
we could go from the exponentiated constraint back to the non-exponentiated ones. 

To see what happens when we project before exponentiation, we go back to (\ref{3.d})
\ba \label{3.p}   
&& <W[\tilde{F}^1,0]\Omega, (-iC^\epsilon_u)\;W[F^1,0]\Omega>
\\
&=&u^{1J}\;f_J(F^1,0)\; <W[\tilde{F}^1,0]\Omega, \partial^\epsilon_1\; W[F^1,0]\Omega>
+u^{2J}\;f_J(F^1,0)\; <W[\tilde{F}^1,0]\Omega, \partial^\epsilon_2\; W[F^1,0]\Omega>
\nonumber
\ea
However, the second term vanishes because it is a linear combination of vectors with
$F^2=\epsilon, \epsilon^2>0$. To compute the next order we introduce the 
translation operators $(T^+_I\; H)[F]=H[F+\epsilon\delta_I],\; 
(T^-_I\; H)[F]=-H[F+\epsilon^2\delta_I]$ to obtain
\be \label{3.q}   
<W[\tilde{F}^1,0]\Omega, (-iC^\epsilon_u)^2\;W[F^1,0]\Omega>
=\epsilon^{-2}\; u^{IJ}\; u^{KL}\sum_{\sigma,\sigma'=\pm} \;
<W[\tilde{F}^1,0],\; (f_J\; T^\sigma_I\; f_L\; T^{\sigma'}_K\; W)[F^1,0]\Omega>
\ee
We see that the terms with $I=2$ and/or $K=2$ respectively produce Weyl elements 
with $F_2=\epsilon, \epsilon^2, 2\epsilon, 2\epsilon^2, \epsilon+\epsilon^2$ respectively 
whose contributions therefore all vanish which leaves us with 
\be \label{3.r}   
<W[\tilde{F}^1,0]\Omega, (-iC^\epsilon_u)^2\;W[F^1,0]\Omega>
=u^{1J}\; u^{1L}
<W[\tilde{F}^1,0],\; (f_J\; \partial^\epsilon_1\; f_L\; \partial_1^\epsilon\; W)[F^1,0]\Omega>
\ee  
Iterating we see that while $[-i C^\epsilon_u]^N$ produces terms with 
$F^2=k \epsilon+l\epsilon^2$ with $k,l\ge 0,\; 0\le k+l \le N$ these all drop out unless 
$k=l=0$. Thus 
only discrete derivatives with respect to the $F^1$ dependence are left over. One 
can therefore set $F^2=0$ also before evaluating the matrix element and we find 
\be \label{3.s}
<W[\tilde{F}^1,0]\Omega, e^{-iC^\epsilon_u}\;W[F^1,0]\Omega>
=<W[\tilde{F}^1,0]\Omega, W[(e^{X^\epsilon_{1,u}}\; K_1)[F^1],0]\Omega> 
=<W_1[\tilde{F}^1]\Omega_1, W_1[(e^{X^\epsilon_{1,u}} \;K_1)[F^1]]\Omega_1> 
\ee
Taking now $\epsilon\to 0$ in the sense described above, we see 
that in this case the quantisation (\ref{3.k}) of the classical truncation
(\ref{3.j}) in fact agrees with the matrix elements of the full theory blocked
from the continuum and the renormalisation flow is already at the fixed point. 

If we enlarge the system to say three canonical pairs and consider subspaces 
${\cal H}_1\subset {\cal H}_{12}\subset {\cal H}$ generated by Weyl elements with
$F_2=F_3=0$ and $F_3=0$ respectively we see that by the same mechanism we can 
project the exponentiated constraints either directly from $\cal H$ to ${\cal H}_1$ or 
first to ${\cal H}_{12}$ and then to ${\cal H}_1$, that is, also the entire family is 
consistent. 

Note that it was crucial in this analysis to have defined the discrete 
derivative in terms of strictly positive shifts by $\epsilon,\epsilon^2$ respectively.
If we had used shifts by $\epsilon, 0$ (forward derivative), $(0,-\epsilon)$ (backward 
derivative) or $\epsilon, -\epsilon$ (antisymmetric derivative) then at various stages 
of the iteration we would have encountered also zero shift in the $F^2$ direction which 
would have produced non-vanishing contributions to the projected matrix elements and due 
to the division by powers of $\epsilon$ would have produced ill-defined results  
after exponentiation in the limit $\epsilon\to 0$.    
      
\subsubsection{Renormalisation flow in terms of projections for general weight $w$}
\label{s3.1.1} 

We proceed analogously to the toy model and write
write $A_{3M}=A_M+[A_{3M}-A_M]=P_M\; A+P_{M3M}^\perp \; A=A_M+A_{M3M}$ where $A_M,\; A_{M3M}$ play
respectively the roles of $A_1,A_2$ in the toy model. Then we regularise the constraint $D^\epsilon_{3M}[u]$ 
and consider its matrix element on the subspace defined by the span of the 
$W_M[F_M]=W_{3M}[F_M]$ given by 
\ba \label{3.15}
&& <\rho_{3M}(W_{3M}[F'_M])\Omega_{3M},\; \rho_{3M}(D^\epsilon_{3M}[u])
\rho_{3M}(W_{3M}[F_M])\Omega_{3M}>_{{\cal H}_{3M}}
\\
&=& <\rho_{3M}(W_{3M}[F'_M])\Omega_{3M},\; 
[i\int\; d^3x\; D^a_j(u, F_M)]\;\{
\delta^\epsilon_M(x)^j_a+ \delta^\epsilon_{M3M}(x)^j_a\}\;
\rho_{3M}(W_{3M}[F_M])\Omega_{3M}>_{{\cal H}_{3M}}
\nonumber
\ea
where 
\ba \label{3.16}
&& \delta^\epsilon_M(x)^j_a\;\rho_{3M}(W_{3M}[F_{3M}])\Omega_{3M}
\nonumber\\
&:=&\frac{1}{\epsilon}
[\rho_{3M}(W_{3M}[F_{3M}+\epsilon\;\Delta_a^j\; P_M(x,.)\; ])\Omega_{3M}
-\rho_{3M}(W_{3M}[F_{3M}+\epsilon^2\;\Delta_a^j\; P_M(x,.)\; ])\Omega_{3M}
\nonumber\\
&&\delta^\epsilon_{M3M}(x)^j_a\;\rho_{3M}(W_{3M}[F_{3M}])\Omega_{3M}
\nonumber\\
&:=& \frac{1}{\epsilon}
[\rho_{3M}(W_{3M}[F_{3M}+\epsilon\;\Delta_a^j\; P_{M3M}(x,.)\; ])\Omega_{3M}
-\rho_{3M}(W_{3M}[F_{3M}+\epsilon^2\;\Delta_a^j\; P_{M3M}(x,.)\; ])\Omega_{3M}
\ea
and where the vector $\Delta_a^j$ has components $(\Delta_a^j)^b_k=\delta_a^b\;\delta^j_k$.
These two operators play the roles of $\partial^\epsilon_1,\partial^\epsilon_2$ in the 
toy model. They perform $\epsilon$ dependent shifts of $F_{3M}$ into the direction 
of the projections $P_M,\; P_{M3M}$ respectively and generate the terms
$\epsilon^n A_{M,a}^j(x),\; \epsilon^n A_{M3M,a}^j(x);\;n=1,2$ in addition to 
$<F_{3M},A_{3M}>_{L_{3M}}$
in the exponent of $W_{3M}[F_{3M}]$ where $F_M$ is considered as a special element 
of $L_{3M}$. Note that $P_M(x,.)\in L_M$ for every fixed $x$ and similar for $P_{M3M}(x,.)$. 
It is easy to see that on functions that are functionally differentiable 
the combination $\delta^\epsilon_M(x)^j_a+\delta^\epsilon_{M3M}(x)^j_a$ reduces to 
$\frac{\delta}{\delta F^a_{3M,j}(x)}$ in the limit $\epsilon\to 0$. 

The mechanism is now completely analogous to the toy model: The N-th power of 
$\rho(D^\epsilon[u])$ produces terms with shifts of 
the form $[k\epsilon + l \epsilon^2] P_{M3M}(x,.)$ with $k,l\ge 0, 0\le k+l\le N$ but only the 
terms with $k=l=0$ survive the matrix element calculation. Writing this out in detail
is a tedious exercise left to the interested reader. It follows that 
\ba \label{3.17}
&& <\rho_{3M}(W_{3M}[F'_M])\Omega_{3M},\; [-i\rho_{3M}(D^\epsilon_{3M}[u])]^N
\rho_{3M}(W_{3M}[F_M])\Omega_{3M}>_{{\cal H}_{3M}}
\nonumber\\
&=& <\rho_{3M}(W_{3M}[F'_M])\Omega_{3M},\; [X^{\epsilon,M}_u]^N
\rho_{3M}(W_{3M}[F_M])\Omega_{3M}>_{{\cal H}_{3M}}
\ea
where for any functional $K[F_M]$ we define 
\be \label{3.17a} 
[X^{\epsilon, M}_u\; K][F_M]:=
[\int\; d^3x\; D^a_j(u, F_M)]\;\
[\delta^\epsilon_M(x)^j_a\;K][F_M]
\ee
It follows that 
\ba \label{3.17b}
&&<\rho_{3M}(W_{3M}[F'_M])\Omega_{3M},\; \rho_{3M}(e^{-i D^\epsilon_{3M,u}})\; 
\rho_{3M}(W_{3M}[F_M])\Omega_{3M}>_{{\cal H}_{3M}}
\nonumber\\
&=&
<\rho_{3M}(W_{3M}[F'_M])\Omega_{3M},\; 
\rho_{3M}(W_{3M}[(e^{X^{\epsilon,M}_u} K_M)[F_M])\Omega_{3M}>_{{\cal H}_{3M}}
\ea
of which we now can take the formal limit $\epsilon\to 0$ on the space of semaring functions 
in $L_M$. This limit coincides with (\ref{3.7}), hence the flow is already fixed pointed.

Finally, note that $(e^{X^M_u} K_M)[F_M]\in L_M$ while the functions 
$D^a_j(u,F_M;.)$ do not belong to $L_M$ in general. This is because 
\be \label{3.17c}
[X^M_u\; K_{M,j}^a(x)][F_M]:=
\int\; d^3y\; D^a_j(u, F_M;y)]\;P_M(y,x)
\ee
is a projected function and further actions of $X^M_u$ do not change this because 
they act on the $F_M$ dependence of (\ref{3.17c}) and never cancel the final projection 
$P_M(x,y)$ no matter how non-linear the function $D^a_j(u,F_M,x)$ may be with respect to 
its $F_M$ dependence. This is the direct analog of the fact that in 
the toy model calculation the contributions from $u^{2I}$ drop out of the matrix element of any 
power of $X_u$ within projected matrix elements. While for the diffeomorphism constraint 
that dependence is in fact linear, for the Hamiltonian constraint it is not, in particular 
for $w-2$ not a non-negative integer multiple of 4. Yet, $[e^{X^M_N} K_M](0,F_M)\in L_M$.
Hence all that was said for the spatial diffeomorphism constraint extends to the Hamiltonian
constraint for any density weight.

\subsection{Renormalisation flow for discretised constraints}
\label{s3.2}

The purpose of this subsection is twofold.  
First, instead of projecting onto the \( L_M \) subspaces, 
we will reformulate the results of the previous subsection in terms of discretised fields in order 
to display renormalisation in terms of the more familiar ``block spin transformations''. 
Second, when discretising the theory on a lattice, multiple discretisation choices, denoted by \( c_M \) in the previous section,
suggest themselves and thus comparing the renormalisation flow starting from different initial discretisation choices is 
of interest. \\
\\
We follow the general framework of appendix \ref{sa}. There we work with a concrete choice 
of embedding map $I_M: l_M\to L_M$ for functions defined on the lattice $\mathbb{N}_M^3$ 
with scalar product $<f_M,g_M>=M^{-3}\;\sum_{m\in \mathbb{N}_M^3} f^\ast_M(m) g_M(m)$ to 
the subspace $L_M$ of the space $L=L_2(T^3,d^3x)$ which is the span of the functions 
$e^{2\pi i n x},\; n\in \mathbb{Z}_M^3$ where the resolution $M$ takes values in the set 
of odd naturals. The chosen map $I_M$ uses the Dirichlet kernel, however, much of what follows 
does not exploit the details of that map, what is important is that $I_M$ has a smooth image
and that its adjoint $I_M^\dagger: L_M\to l_M$ has the following properties: 1. $I_M$ is an 
isometric embedding and 2. $P_M=I_M\cdot I_M^\dagger$ is the orthogonal projection $L\to L_M$. 
The finer details of $I_M$ are only important when trying to interpret discretised functions 
$f_M=I_M^\dagger F$ as related to the restrictions $\bar{F}_M$ of $F$ to the lattice points 
$x^M_m=m/M,\; m\in  \mathbb{N}_M^3$. 
Here we deviate from the notation in appendix \ref{sa} 
and denote here by $I_M^\dagger$ what is called $I^\ast_M$ there in order to avoid confusion
with complex conjugation.   
        
The discretisation of the fields with values in $l_M$ follows the pattern of appendix \ref{sa} and is given in terms of the 
continuum fields in $L$ by
\begin{align}
\label{adjoint}
    (e^a_{M,j})(m):=(I^\dagger_M E^a_{j})(m):=\braket{\chi^M_m,E^a_j}_{L_M},\;\;
    (a_{M,a}^j)(m):=(I^\dagger_M A_{a}^j)(m):=\braket{\chi^M_m,A^j_{a}}_{L_M},
\end{align}
where \(m \in\mathbb{N}^3_M\). 
Moreover, the injection into $L_M$ is defined by
\begin{align}
\label{injection}
    (I_M e^a_{M,j})(x):=\frac{1}{M^3}\sum_{m\in\mathbb{Z}_M^3} e^a_{M,j}(m)\chi^M_m(x),\;\;
    (I_M a_{M,a}^j)(x):=\frac{1}{M^3}\sum_{m\in\mathbb{Z}_M^3} a_{M,a}^j(m)\chi^M_m(x).
\end{align}
This relates the discretised fields in $l_M$ to the projected fields
$A_{M,a}^j(x),\;E^a_{M,j}(x)$ in $L_M$ via $A^j_{M,a}=I_M \cdot a^j_{M,a}, E^a_{M,j}=I_M\cdot e^a_{M,j}$
Substituting this in the previous expressions results in a flow strictly equivalent to the 
previous projection formalism but it maybe favoured by those who prefer displaying 
the constraints as real space discretisations rather than subspace projections. 
The resulting expressions in fact motivate different discretisations 
which look more local. 
The corresponding flow then converges more slowly to the 
fixed point and thus provides a further test of the renormalisation method proposed 
in this series of works.  

In order to be able to perform the integrals over $x$ explicitly and to display 
the constraints as explicitly as possible in a strictly discretised form, we focus on the case that 
$w-2=4k,\; k\in \mathbb{N}_0$ so that the Hamiltonian constraint is a homogeneous polynomial of degree 
$3+6k$ with one factor of $a_M$ and $2+6k$ factors of $e_M$. Indeed for other values of $w$
one could not do the $x$ integral easily because then we would need to control the 
integral of 
\be \label{3.20}
|\det(E_M(x))|^{(w-2)/2}=|\frac{1}{3!\; M^9}\epsilon_{abc}\epsilon^{jkl}\sum_{m_1, m_2, m_3} 
\;\prod_{I=1}^3\;\chi^M_{m_I}(x)\; e^a_{M,j}(m_1) \; e^b_{M,k}(m_2) \; e^c_{M,l}(m_3)|^{(w-2)/2}
\ee
against three factors of $\chi^M_m$ functions which displays fractional (inverse) powers of the $\chi^M_m$. 
This introduces a high degree of non-locality as we will discuss in more detail in the final 
subsection of this section. This is the reason why 
one typically considers more local initial discretisations as starting points of the renormalisation flow, 
justified by the fact that $\chi^M_m(x)$ turns into $\delta(x,y)$ when 
we take simultaneously $M\to\infty,\; m\to \infty$ such that $x^M_m=\frac{m}{M}\to y$. 

Since the essential features become already apparent for the simplest case $k=0$ we focus 
on that in what follows. Then we have explicitly with the notation $<f>:=\int\; d^3x \; f(x)$
the exact expressions strictly equal to the ones of the previous section 
\ba \label{3.21}
C_M[r] &=& -\frac{1}{M^3}\sum_m\; <r^j_{,a}\;\chi^M_m>\; e^a_{M,j}(m) 
\nonumber\\
D_M[u] &=& \frac{1}{M^6}\sum_{m_1,m_2}\; <[u^a_{,b}\chi^M_{m_2}-(\delta^a_b\; u^c\; \chi^M_{m_2})_{,c}]\;\chi^M_{m_1}> 
a^j_{M,a}(m_1)\; e^b_{M,j}(m_2)
\nonumber\\
H_M[N] &=& \frac{1}{M^9}\epsilon_{jkl}\;\sum_{m_1,m_2,m_3}\; <(N\;\chi^M_{m_2}\; \chi^M_{m_3})_{,b}\;\chi^M_{m_1}>\;
a^j_{M,a}(m_1)\;e^{[a}_{M,k}(m_2)\; e^{b]}_{M,l}(m_3)  
\ea
We notice that these constraints, even though polynomial, display an inherent {\it spatial non-locality} 
because we use the Dirichlet rather than the Haar kernel. For the Haar kernel we would have 
\be \label{3.22}
\prod_{I=1}^N \chi^M_{m_I}=\chi^M_{m_N}\;\prod_{I=1}^{N-1}\; \delta_{m_I,m_N}\
\ee
which would let the double and triple sums immediately collapse to a single sum, at the price 
that the derivatives that appear in (\ref{3.21}) become distributional. On the other hand these 
multiple sums are also not totally non-local but rather quasi-local. This is because upon 
Fourier expanding $f=\sum_{n\in \mathbb{Z}^3} \hat{f}(n)\; e_n$ we have for instance for simplicity in one dimension
\begin{align} \label{3.23}
 <f\; \chi^M_{m_1}\; \chi^M_{m_2}>=\sum_{n_1,n_2\in \mathbb{Z}_M^3}\; \hat{f}(n_1+n_2)\; 
e^M_{n_1}(m_1)e^M_{n_2}(m_2)=\sum_{\substack{|n|\le M-1 \\n_2, \;n-n_2 \in \mathbb{Z}_M^3}}\; e^M_{n}(m_1) \hat{f}_M(n)\;e^M_{n_2}(m_2-m_1).
\end{align}
Note that at given $n$ the summation range of $n_2$ is 
max$(-(M-1)/2, -(M-1)/2+n)\le n_2 \le$min$((M-1)/2,n+(M-1)/2)$. 
If $f$ has compact momentum support say in $|n|\le k$ then for $M\gg k$ the range of 
$n_2$ becomes almost independent of $n$ and the expression (\ref{3.23}) approximately 
factorises
\be \label{3.24}
<f\; \chi^M_{m_1}\; \chi^M_{m_2}>\approx 
f(x^M_{m_1})\; \chi^M(x^M_{m_2}-\chi^M_{m_1})=M\; \delta_{m_1,m_2}\; f(x^M_{m_1})
\ee 
where $x^M_m:=\frac{m}{M}$ and the properties of the Dirichlet kernel were used. Therefore the multiple sums are 
quasi-local in the limit $M\to \infty$.

These considerations motivate to propose a different initial discretisation of the constraints 
which we call ``local'' and that appear to be ``more natural''. 
To that end we define the discrete derivative $\partial_{M,a}=I_M^\dagger\partial_a I_M$
and have since $\partial_a$ preserves $L_M$ 
\be \label{3.25}
\partial_b E^a_{M,j}=P_M\;\partial_b E^a_{M,j}=I_M\;[I_M^\dagger \partial_b I_M] e^a_{M,j}=
I_M\; \partial_{M,b} e^a_{M,j}
\ee
Hence we have the still exact relations
\ba \label{3.26}
&& D^a_j(u,E_M)=[u^a_{,b}-\delta^a_b \; u^c_{,c}]\; [I_M e^b_{M,j}]-u^b\; [I_M \partial_{M,b} e^a_{M,j}]
\\
&& H^a_j(N,E_M)=\epsilon_{jkl} \{N_{,b} [I_M e^a_{M,k}]\; [I_M e^b_{M,k}]
+ N( [I_M \partial_{M,b} e^a_{M,k}]\; [I_M e^b_{M,k}]
+[I_M e^a_{M,k}]\; [I_M \partial_{M,b} e^b_{M,k}])\}
\nonumber
\ea
Performing the $x$ integral over (\ref{3.26}) against $A_{M,a}^j(x)$ now still results in 
double and triple sums respectively with "couplings"
\begin{align} \label{3.27}
    u^a_{M,b}(m_1,m_2)&:=<[u^a_{,b}-\delta^a_b \; u^c_{,c}]\chi^M_{m_1}\chi^M_{m_2}>, \;\;\;
    u^{M,b}(m_1,m_2):=<u^b\chi^M_{m_1}\chi^M_{m_2}>,\nonumber \\
    N_{M,b}(m_1,m_2,m_3)&:=<N_{,b}\;\chi^M_{m_1}\chi^M_{m_2}\chi^M_{m_3}>, \;\;\;
N_M(m_1,m_2,m_3):=<N\;\chi^M_{m_1}\chi^M_{m_2}\chi^M_{m_3}>.
\end{align}
Note that the couplings, depending linearly on the continuous functions $u,N$, can be interpreted 
as an ``automatically discretisation'' of $u,N$, although depending on more than one lattice point. 
In the limit of large $M$ these expressions become rather concentrated at a single lattice point.
In that sense $u,N$ need not to be discretised by hand, it happens automatically.

Then the above arguments motivates to consider instead of these exact expressions
as a starting point of the flow for instance 
the expressions
\begin{align} \label{3.28}
C^{{\sf loc}}_M[r] &:= -\frac{1}{M^3}\sum_m\; r^j_{,a}(x^M_m)\; e^a_{M,j}(m) 
\nonumber\\
D^{{\sf loc}}_M[u] &=\frac{1}{M^3}\sum_M\; a^j_{M,a}(m)\;\{[u^a_{,b}-\delta^a_b\; u^c_{,c}](x^M_m)\;e^b_{M,j}(m)
-u^b(x^M_m)\;\partial_{M,b} e^a_{M,j}(m)
\nonumber\\
H^{{\sf loc}}_M[N] &= \frac{1}{M^3}\epsilon_{jkl}\;\sum_m\; a^j_{M,a}(m) \times
\nonumber\\
& \{N_{,b}(x^M_m)\; e^a_{M,k}(m)\; e^b_{M,k}(m)
+ N(x^M_m)\;( [\partial_{M,b} e^a_{M,k}](m)\; e^b_{M,k}(m)
+e^a_{M,k}(m)\; [\partial_{M,b} e^b_{M,k}](m))\}
\end{align}
where $\partial_{M,b}$ could now mean one of the ``standard'' discrete derivatives such as 
the forward derivative $[\partial_{M,b}f_{M}](m)=M[f_M(m+\delta_b)-f_M(m)$ with the lattice 
vector with components $[\delta_b]^a=\delta_b^a$. 

Now we have shown in the previous section that (\ref{3.21}) in its exponentiated form is in 
fact a fixed point of the flow including the essential, mere quasi-locality displayed in
(\ref{3.27}). Therefore 
the exponentiation of (\ref{3.28}) is not a fixed point and the flow is non-trivial. 
In the following we will derive and exemplify the renormalisation group equations for Hamiltonian 
and spatial diffeomorphism constraints in terms of ``coupling parameters''
$u^{a(r)}_{M,b}(m_1,m_2), \; N^{(r)}_{M,b}(m_1,m_2,m_3)$ where $r$ denotes the iteration step.
We will show that (\ref{3.27}) is a fixed point of these equations and that the iteration 
starting with the initial values 
$u^{a(0)}_{M,b}(m_1,m_2), \; N^{(0)}_{M,b}(m_1,m_2,m_3)$ displayed in (\ref{3.28}) converge 
to (\ref{3.27}). \\
\\
To do this, we adopt the following strategy:
\begin{itemize}
\item[i.]
To avoid confusion we note the following:\\
In \cite{TZ} the option was considered to discretise not only the dynamical fields (here $A,E$) but 
also the smearing functions (here $r,u,N$) ``by hand'' in addition to the automatic 
discretisation mentioned above. Then the r-th renormalisation step consists in computing say 
$D^{(r+1)}_M[u_M]$ on ${\cal H}_M$ by projecting $D^{(r)}(I_{M3M} u_{3M})$ on ${\cal H}_{3M}$ to the subspace 
${\cal H}_M$. There are arguments in favour of and against coarse graining also the smearing functions and not only the canonical fields.
The pro argument is that the smearing functions are in principle also canonical fields (such as here lapse and shift), 
it is just that they are considered pure gauge Lagrange multipliers 
as dictated by the primary constraints that their conjugate momenta
have to vanish. Thus, coarse graining also the smearing fields would put all fields on equal footing.
The contra argument is that at a fixed smearing field we obtain a constraint that has the same status as 
a Hamiltonian in an unconstrained QFT with some background structure (say a self-interacting scalar field in 
a background spacetime) where here the background structure is given by the smearing field. As one certainly 
would not coarse grain the background metric when renormalising QFT in a background spacetime, one could 
argue that one should not coarse grain the smearing fields. In \cite{TZ} it was shown that 
both flow options have the same fixed point. In what follows we will not discretise the smearing fields
``by hand'' because this would just blow up the formalism and does not improve the convergence rate of the flow.\\  
\item[ii.]
The subtleties related to the discontinuity of the Narnhofer-Thirring representation of course 
transfer to the present subsection as well. We may reformulate these as follows, presenting the 
issue from a  different, topological angle:\\ 
In the previous subsection we have demonstrated that due to the discontinuity of the Narnhofer-Thirring
representation not only the renormalisation flow must be formulated for the exponentiated constraints but
also that the flow has to be formulated using the the discrete topology rather than 
relying on continuity with respect to the weak operator topology. We have seen e.g. 
that one cannot just take the flow 
for a generic label of the Weyl elements and then take a limit to a restricted class of labels as 
if that limit would be well defined in the weak operator topology. To make this precise and to 
formulate the flow in terms of couplings, consider the set of couplings 
$G$ with one of its standard function topologies, a set of functions $L$ on $G$ with one of the 
standard topologies (such as the functions or Hamiltonian vector fields of those functions 
on phase space with topology inherited from that of the 
phase space) and a set of functionals $E$ on $L$ (such as vacuum expectation value functionals with respect 
to Weyl elements labelled by $F\in L$). Given $F\in L$ and 
$e\in E$ we obtain a function $e_F:=e\circ F$ on $G$. Now the topologies on $G,L$ are such that the elements 
$F\in L$ are continuous functions on $G$. But $e_F$ is not continuous with respect to the given 
topology on $G$ but rather with respect to the discrete topology on $G$:     
In the discrete topology,   
every coupling $g$ defines an open set with one element $\{g\}$, 
hence the smallest open neighbourhood of $g$ that contains $g$ is $\{g\}$.
Then due to discontinuity of a typical element $e_F$ in the given topology of $G$, the only 
open neighbourhood $O_\epsilon$ in the discrete topology of a coupling $g_0$ 
such that $|e_F(g)-e_F(g_0)|<\epsilon$ 
for any sufficiently small $\epsilon$ is given by $O_\epsilon=\{g_0\}$. For instance
$e_F(g)=\delta_{g,g_0}$ is a typical example. Thus $e_F$ is continuous in the 
discrete topology in the mathematical sense but taking limits of sequences $n\mapsto g_n$ becomes trivial, one must in fact 
take $N(\epsilon)=\infty$ in order that $|e_F(g_n)-e_F(g_0)|<\epsilon$ for all $n>N(\epsilon)$
for a typical sequence that converges in the given topology of $G$ such as $g_n=(1+1/n) g_0$. 
In other words, the only such sequences with respect to which the $e_F$ are continuous are those 
which eventually become constant $g_n=g_0$ for all $n>N_0$. Therefore taking limits for the 
$e_F$ is the same thing as evaluating at the limit, there is no non-trivial limiting process possible.

We are therefore forced to proceed as follows: To derive the renormalisation 
flow as a map on the space $G$, we use first the same regularisation procedure for the exponentiated 
constraints such as $e^{-i D_M[u]}$ as in the previous subsection. 
The action on and matrix elements between vectors $W_M[F_M]\Omega_M$
of the regulated object 
$e^{-i D^\epsilon_M[u]}$ can be computed and amounts to 
a map $F_M\mapsto (e^{X^\epsilon_u}\cdot K_M)[F_M]$ on the space $L_M$. This expression allows a limit $\epsilon\to 0$
in the topology of $L_M$ denoted by $(e^{X_u}\cdot K_M)[F_M]$. And taking the action at that 
limit point corresponds to taking the limit in the discrete topology of $L_M$ and defines what we 
mean by $e^{-i D_M[u]}$. However, for the same reason,
we cannot take $e^{-i D_{3M}[u]}$ as derived 
in terms of its matrix elements of generic vectors $W_{3M}[F_{3M}]\Omega_{3M}$ and then take a limit 
$F_{3M}\to F_M$, rather we must compute the matrix elements directly between the vectors 
$W_{3M}[F_{M}]\Omega_{3M}$. Having obtained that object we find that it corresponds to a 
another map $F_M\mapsto (e^{\tilde{X}^M_u}\cdot K_M)[F_M]$. The flow is therefore a flow 
of Hamiltonian vector fields and thus indirectly a flow of couplings because the Hamiltonian 
vector fields are parametrised by those. The difference with the previous subsection is that we consider 
generic couplings and use the equivalent formulation in terms of the discretised functions 
$f_M=I_M^\dagger\cdot F_M$ rather than the projected functions $F_M=I_M\cdot f_M$.
\end{itemize}
According to this strategy, we write the classical discretised constraints in the 
general form 
\ba \label{3.29}
D_M[u]=M^{-6}\sum_{m_1,m_2\in \mathbb{N}_M^3}\;u^a_{M,b}(m_1,m_2) \; a^j_{M,a}(m_1)\; e^b_{M,j}(m_2)
\nonumber\\
H_M[N]=M^{-9}\sum_{m_1,m_2,m_3\in \mathbb{N}_M^3}\;N_{M,b}(m_1,m_2,m_3) \; 
\epsilon_{jkl}\; a^j_{M,a}(m_1)\; e^{[a}_{M,k}(m_2)\; e^{b]}_{M,l}(m_3) 
\ea
in which we leave the form of the functions $u^a_{M,b},\;N_{M,b}$ unspecified.
The non-trivial classical Poisson brackets are  
\be \label{3.30}
\{e^a_{M,j}(m_1),a^k_{M,b}(m_2)\}=[M^3\; \delta_{m_1,m_2}]\;\delta^a_b\; \delta^k_j=:
\delta_M(m_1,m_2)\;\delta^a_b\; \delta^k_j
\ee
We have the following identity on Weyl elements 
\ba \label{3.31}
W_M[F_M] &=& e^{-i<A_M,F_M>_{L_M}}=
e^{-i<I_M \cdot a_M,I_M\cdot f_M>_{L_M}}= 
e^{-i<a_M,f_M>_{l_M}}=:w_M[f_M]
\nonumber\\
<a_M,f_M> &=& M^{-3}\sum_{m\in \mathbb{N}_M^3} a^j_{M,a}(m)\; f^a_{M,j}(m)
\ea
The operators corresponding to $a^j_{M,a}(m), e^a_{M,j}(m)$ in the Narnhofer-Thirring representation
act on $w_M[f_M]\Omega_M$ formally by multiplication by $a^j_{M,a}(m)$ and by derivation as 
$i\delta/\delta a^j_{M,a}(m)$ where 
\be \label{3.31a}
\frac{\delta}{\delta f_M(m)}:= M^3\frac{\partial}{\partial f_M(m)}
\ee
obeying the canonical quantisation of(\ref{3.30}). The latter operation is well defined and returns 
$f^a_{M,j}(m)$ as eigenvalue while the former operation must be approximated by multiplication 
by $i\epsilon^{-1}[e^{-i\epsilon a^j_{M,a}(m)}-e^{-i\epsilon^2 a^j_{M,a}(m)}]$ causing the 
shift $w_M[f_M]\to i\epsilon^{-1}(w_M[f_M+\epsilon \delta^j_{m,a}]-
w_M[f_M+\epsilon^2\; \delta^j_{m,a}])$ where the discrete distribution has components 
$[\delta^j_{m,a}]^b_k(\tilde{m})=M^3\delta^b_a\delta^j_k\delta_{m,\tilde{m}}$. We will not 
go through all the steps of the previous subsection again but merely write the end result
\ba \label{3.32}
e^{-iD_M[u]}\; w_M[f_M]\Omega_M &=& w_M[(e^{x^M_u}\cdot k_M)[0,f_M]]\Omega_M 
\nonumber\\
e^{-iH_M[u]}\; w_M[f_M]\Omega_M &=& w_M[(e^{x^M_N}\cdot k_M)[0,f_M]]\Omega_M 
\ea
where $k_M=\{k^a_{M,j}[.,.](m )\}$ are the coordinate functionals on the discretised phase space defined by \\
$(k^a_{M,j}[a_M,e_M])(m)=e^a_{M,j}(m)$ and $x^M_u, x^M_N$ are the Hamiltonian vector fields
corresponding to (\ref{3.29}). Explicitly, for any functional $k[f_M]$ they read 
with $\frac{\delta}{\delta f^a_{M,j}(m)}:=M^3\; \frac{\partial}{\partial f^a_{M,j}(m)}$
\ba \label{3.33}
(x^M_u \cdot k)[f_M]:=
M^{-6}\sum_{m_1,m_2\in \mathbb{N}_M^3}\;u^a_{M,b}(m_1,m_2) \; f^b_{M,j}(m_2)\; 
\frac{\delta k[f_M]}{\delta f^a_{M,j}(m_1)}
\nonumber\\
(x^M_N \cdot k)[f_M]=M^{-9}\sum_{m_1,m_2,m_3\in \mathbb{N}_M^3}\;N_{M,b}(m_1,m_2,m_3) \; 
\epsilon_{jkl}\; f^{[a}_{M,k}(m_2)\; f^{b]}_{M,l}(m_3)\; \frac{\delta k[f_M]}{\delta f^a_{M,j}(m_1)}
\ea
We now go through the same analysis but consider instead matrix elements of powers of 
$D_{3M}[u], H_{3M}[N]$ 
between vectors of the form $w_{3M}[I_{M3M}\cdot f_M]$ where $I_{M3M}=I_{3M}^\dagger I_M:\; l_m\to l_{3M}$,
see appendix \ref{sa}. As in the previous section, the operators $e^a_{3M,j}(m')$ are diagonal 
on those vectors with eigenvalues $[I_{M3M}\cdot f^a_{M,j}](m'),\;m'\in \mathbb{N}_{3M}^3$.
As in the previous subsection where $W_{3M}[F_M]$ in fact only depends on $A_M=P_M\cdot A_{3M}$
the Weyl element $w_{3M}[I_{M3M}\cdot f_M]$ does not depend on all components of $a_{3m}$ but 
only on $p_{M3M}\cdot a_{3M}$ where $p_{M3M}=I_{M3M}\cdot I_{M3M}^\dagger$. This is because 
\be \label{3.33a}
I_{M3M}^\dagger\cdot I_{M3M}=I_M^\dagger\cdot I_{3M}\cdot I_{3M}^\dagger\cdot I_M
=I_M^\dagger\cdot P_{3M}\cdot I_M= I_M^\dagger\cdot I_M=1_{l_M}
\ee
as $I_M$ has image in $L_M$ on which $P_M$ projects as well as $p_{M3M}^\dagger=p_{M3M}$ and 
\be \label{3.34}
p_{M3M}^2=[I_{3M}^\dagger\cdot I_M\cdot I_M^\dagger\cdot I_{3M}]^2
=[I_{3M}^\dagger\cdot P_M\cdot I_{3M}]^2=
I_{3M}^\dagger\cdot P_M\cdot P_{3M}\cdot P_M\cdot I_{3M}]^2=p_{M3M}
\ee
as $P_M$ is a subprojection of $P_{3M}$. Hence $p_{M3M}$ is the analog of $P_M$ in the discrete 
setting and we have the identity $p_{M3M}\cdot I_{M3M}=I_{M3M}$ so that 
$<a_{3M},I_{M3M}\cdot f_M>_{l_{3M}}=<p_{M3M}\cdot a_M,I_{M3M}\cdot f_M>_{l_{3M}}$. Therefore as 
in the previous subsection we decompose the operator $a_{3M}$ appearing in $D_{3M}[u],\;H_{3M}[N]$ 
as $p_{M3M}\cdot a_{3M}+p_{M3M}^\perp \cdot a_{3M}$ with $p_{M3M}^\perp=1_{l_{3M}}-p_{M3M}$ and 
replace both terms separately by regulated multiplication operators causing $\epsilon, \epsilon^2$ 
depending shifts in those orthogonal directions in the space of the $f_{3M}$. By the same mechanism 
as in the previous subsection, the $p_{M3M}^\perp$ shifts drop out in matrix elements, leaving us 
formally with the contribution
\ba \label{3.35}
&& [p_{M3M}\cdot a_{3M}](m')\; w_{3M}[I_{M3M}\cdot f_M]=
[I_{M3M}\cdot (I_{M3M}^\dagger\cdot a_{3M}0](m')\; e^{-i<I_{M3M}^\dagger a_{3M},f_M>_{l_M}}
\nonumber\\
&=& i [I_{M3M}\cdot \frac{\delta}{\delta f^a_{M,j}(.)}](m')\; w_{3M}[I_{M3M}\cdot f_M]
\ea
It follows that formally in matrix elements 
\ba \label{3.36}
&& -i D_{3M}[u]\; w_{3M}[I_{M3M}\cdot f_M]
=M^{-6}\sum_{m_1,m_2\in  \mathbb{N}_M^3} \; 
[(I_{M3M}^\dagger \times I_{M3M}^\dagger)\cdot u^a_{3M,b}](m_1,m_2)
\times \nonumber\\
&& \; f^b_{M,j}(m_2) \; \frac{\delta}{\delta f^a_{M,j}(m_1)} w_{3M}[I_{M3M}\cdot f_M]
\nonumber\\
&&-i H_{3M}[u]\; w_{3M}[I_{M3M}\cdot f_M]
=M^{-9}\sum_{m_1,m_2,m_3\in  \mathbb{N}_M^3} \; 
[(I_{M3M}^\dagger \times I_{M3M}^\dagger\times I_{M3M}^\dagger)\cdot N_{3M,b}](m_1,m_2,m_3)\; 
\times \nonumber\\
&& \epsilon_{jkl}
f^{[a}_{M,k}(m_2)\; f^{b]}_{M,l}(m_3)\; \frac{\delta}{\delta f^a_{M,j}(m_1)} w_{3M}[I_{M3M}\cdot f_M]
\ea
Here we have used that e.g. in $D_{3M}[u]$ we encounter sums of the form      
\ba \label{3.37}
&& (3M)^{-3}\sum_{m'_2\in \mathbb{N}_{3M}^3}\; u^a_{3M,b}(m'_1,m'_2) \; (I_{M3M}\cdot f_M)(m'_2)
=<u^a_{3M,b}(m'_1,.),I_{M3M}\cdot f_m>_{l_{3M}}
\\
&=& <I_{M3M}^\dagger\cdot u^a_{3M,b}(m'_1,.),f_m>_{l_{M}} 
=M^{-3}\sum_{m_2\in \mathbb{N}_M^3}\; [(1_{3M}\times I_{M3M}^\dagger0\cdot u^a_{3M,b}](m'_1,m_2) \; 
f_M(m_2)
\nonumber
\ea
Exponentiating we find in matrix elements that 
\be \label{3.37a}
e^{-i D_{3M}[u]}\cdot w_{3M}[I_{M3M}\cdot f_M]=w_{3M}[I_{M3M}\cdot (e^{\tilde{x}^M_u} \cdot k_m)[0,f_M]],\;\;
e^{-i H_{3M}[N]}\cdot w_{3M}[I_{M3M}\cdot f_M]=w_{3M}[I_{M3M}\cdot (e^{\tilde{x}^M_N} \cdot k_m)[0,f_M]]
\ee
where the quantities with the tilde are the {\it renormalised vector fields}
\ba \label{3.38}
(\tilde{x}^M_u \cdot j)[f_M] &:=&
M^{-6}\sum_{m_1,m_2\in \mathbb{N}_M^3}\;
[(I_{M3M}^\dagger \times I_{M3M}^\dagger)\cdot u^a_{3M,b}](m_1,m_2) \; f^b_{M,j}(m_2)\; 
\frac{\delta j[f_M]}{\delta f^a_{M,j}(m_1)}
\\
(\tilde{x}^M_N \cdot j)[f_M] &:=& M^{-9}\sum_{m_1,m_2,m_3\in \mathbb{N}_M^3}\;
[(I_{M3M}^\dagger \times I_{M3M}^\dagger\times I_{M3M}^\dagger)\cdot N]_{3M,b}(m_1,m_2,m_3) \; 
\times\nonumber\\
&& \epsilon_{jkl}\; f^{[a}_{M,k}(m_2)\; f^{b]}_{M,l}(m_3)\; \frac{\delta j[f_M]}{\delta f^a_{M,j}(m_1)}
\nonumber
\ea
acting on functionals $j[f_M]$ such as $k_M[0,f_M]$. Here we have exploited the fact that $\tilde{x}^M_u,\;\tilde{x}^M_N$ 
do not act on $I_{M3M}$ in 
$w_{3M}[I_{M3M}\cdot f_M]=e^{-i<I_{M3M}^\dagger\cdot a_{3M},f_M>_{l_M}}$ so that the mechanism of 
the exponentiation is the same as in (\ref{3.32}) except that $a_M$ is replaced by $I_{M3M}^\dagger a_{3M}$.

The upshot is that renormalisation is now mapped into the space of couplings. Given initial 
data $u^{a(0)}_{M,b}:\; l_M^2\to \mathbb{R},\; N^{(0)}_{M,b}:\;l_M^3\to \mathbb{R}$ we obtain the 
flow equations
\be \label{3.39}
u^{a(r+1)}_{M,b}=(I_{M3M}^\dagger \times I_{M3M}^\dagger)\cdot u^{a(r)}_{3M,b},\;
N^{(r+1)}_{M,b}=(I_{M3M}^\dagger \times I_{M3M}^\dagger \times I_{M3M}^\dagger)\cdot N^{(r)}_{3M,b}
\ee
It is easy to check that (\ref{3.27}) corresponding to 
\be \label{3.40}
u^a_{M,b}(m_1,m_2)=<\chi^M_{m_1}[u^a_{,b}\chi^M_{m_2}-\delta^a_b\;(u^c\;\chi^M_{m_2})_{,c}]>,\;
N_{M,b}(m_1,m_2,m_3)=<\chi^M_{m_1}\;(N\chi^M_{m_2}\;\chi^M_{m_3})_{,b}>
\ee
is a fixed point of (\ref{3.39}). This is a consequence of the fact that 
\be \label{3.41}
[I_{M3M}^\dagger \chi^{3M}_\cdot](m,x)=<\chi^M_m, I_{3M}\cdot \chi^{3M}_\cdot(x)>_{L_{3M}}
=(3M)^{-3}\sum_{m'\in \mathbb{N}_{3M}^3}\; <\chi^M_m,\chi^{3M}_{m'}>\;\chi^{3M}_{m'}(x)  
=(P_{3M}\cdot \chi^M_m)(x)=\chi^M_m(x)    
\ee
where the fact that the functions $\chi^M_m$ are real valued, the completeness relation 
$\sum_{m'\in \mathbb{N}_{3M}^3}\; \chi^{3M}_{m'}(x)\; \chi^{3M}_{m'}(y)=(3M)^3\; P_{3M}(x,y)$ 
and $\chi^M_m\in L_M\subset L_{3M}$ was used. 

Finally note that the discretised fixed point family of Hamiltonian vector fields $x^M_u, \; x^M_N$ 
precisely corresponds to the projected fixed point family $X^M_u, \; X^M_N$. To see this we write 
their actions on functionals $J,j$, e.g. 
\ba \label{3.41a}
&& (X^M_u\cdot J)[F_M] = \int\; d^3x\int\; d^3y\; U^a_{M,b}(x,y)\; F^b_{M,j}(y)\; \frac{\delta J[F_M]}{\delta F^a_{j,M}(x)}
\\
&& (x^M_u\cdot j)[f_M] = M^{-6}\;\sum_{m_1,m_2}\; u^a_{M,b}(m_1,m_2)\; f^b_{M,j}(m_2)\; \frac{\delta j[f_M]}{\delta f^a_{j,M}(m_1)}   
\nonumber\\
&& U^a_{M,b}(x,y)=\int\; d^3z\; P_M(x,z)\; [u^a_{,b}(z)\;P_M(y,z)-\delta^a_b\;(u^c P_M(y,.))_{,z^c}(z)]
\nonumber\\
&& u^a_{M,b}(m_1,m_2)=\int\; d^3z\; \chi^M_{m_1}(z)\; [u^a_{,b}(z)\chi^M_{m_2}(z)-\delta^a_b\;(u^c \chi^M_{m_2})_{,z^c}(z)]
\nonumber
\ea
and note the relation $j[f_M]=J[I_M\cdot f_M]$. Thus by the chain rule 
\ba \label{3.41b}
\frac{\delta j[f_M]}{\delta f^a_{j,M}(m)} 
&=& \int\; d^3x\; [\frac{\delta J[F]}{\delta F^b_k(x)}]_{F=F_M}\;\frac{\delta (I_M\cdot f^b_{M,k})(x)}{\delta f^a_{j,M}(m)} 
\nonumber\\
&=& \int\; d^3x\; [\frac{\delta J[F]}{\delta F^b_k(x)}]_{F=F_M}\;M^3\;[M^{-3}\;\sum_{\tilde{m}} \chi^M_{\tilde{m}}(x)\;
\frac{\partial f^b_{M,k})(\tilde{m})}{\partial f^a_{j,M}(m)}] 
\nonumber\\
&=& \int\; d^3x\; \chi^M_{m}(x)\;[\frac{\delta J[F]}{\delta F^a_j(x)}]_{F=F_M}
=(I_M^\dagger \cdot \frac{\delta J[F]}{\delta F^a_j(.)}]_{F=F_M})(m)
\ea
It follows the identity
\be \label{3.41c}
I_M\cdot \frac{\delta}{\delta f_M}=P_M\;[\frac{\delta}{\delta F}]_{F=F_M}=:\frac{\delta}{\delta F_M}
\ee
on functionals $J$ of $F_M=I_M\cdot f_M$, see (\ref{3.8b}). We use this identity in (\ref{3.41a}) and find 
\be \label{3.41d}
(X^M_u\cdot J)[F_M]=M^{-6}\sum_{m_1,m_2} \; [(I_M^\dagger\times I_M^\dagger)\cdot U^a_{M,b}](m_1,m_2) \;
f^b_{M,j}(m_2)\; \frac{\delta j[f_M]}{\delta f^a_{j,M}(m_1)}
\ee
Finally we note $(I_M^\dagger\cdot P_M)(m,x)=<\chi^M_m,P(.,x)>=\chi^M_m(x)$ to see that 
$(X^M_u\cdot J)[F_M]=(x^M_u\cdot j)[f_M]$. The considerations for $X^M_N,\; x^M_N$ are analogous.\\ 
\\
\\
\\
In the rest of this section we study the flow in the discretised language using the "localised"
form of the constraints (\ref{3.28}). It will be sufficient to do this for the Hamiltonian 
constraint as its treatment includes all the technical steps required for the spatial
diffeomorphism constraint. Moreover, we will work with the antisymmetric lattice derivative 
$[\partial_{Mb}\;f_M](m):=\frac{M}{2}\;[f_M(m+\delta_b)-f_M(m-\delta_b)],\;(\delta_b)^a=\delta^a_b$
which simplifies the notation as summation by parts does not produce a new adjoint lattice 
derivatives but just its negative. We may then cast (\ref{3.28}) into the form
\ba \label{3.50}
H^{{\sf loc}}_M[N] &=& \frac{1}{M^9}\epsilon_{jkl}\;\sum_{m\in \mathbb{N}_M^9}\; a^j_{M,a}(m_1)\; 
e^a_{M,k}(m_2)\; e^b_{M,l}(m_3)\; N^{(0)}_{Mb;m_1,m_2,m_3}
\\
N^{(0)}_{Mb;m_1,m_2,m_3} &=& M^6\;[N_{,b}(x^M_{m_1})\delta_{m_1,m_2}\delta_{m_1,m_2}
-N(x^M_{m_1})\;([\partial_{Mb}\delta_{m_1}](m_2)\;\delta_{m_1,m_3}
\delta_{m_1,m_2}[\partial_{Mb}\delta_{m_1}](m_3))]
\nonumber
\ea
where we wrote $\delta_{m_1}(m_2):=\delta_{m_1,m_2}$ to define a function of $m_2$ parametrised 
by $m_1$. This form of writing the initial value of the flow is convenient as it transparently maps the 
flow entirely on the coupling function $N_{Mb;m_1,m_2,m_3}$. This is the same strategy followed 
in \cite{RZ-T} for P$_2(\Phi)$ theory but here we encounter more complications because 
the lapse function is not a constant and the coupling depends on discrete derivatives. 
Note also that $N_{,b}(x^M_m)$ denotes the continuum derivative of the function evaluated at the lattice 
point $x^M_m$ and not the discrete derivative of the function restricted to the lattice
$[\partial_{Mb} N_M](m),\; N_M(m):=N(x^M_m)$ which also would be a natural initial value. That 
choice leads to no new effects as compared to what we study below and we therefore refrain from 
that option.

The aim will be to show that the flow produces a sequence $N^{(r)}_{Mb;m_1,m_2,m_3}$ which converges,
in a sense to be specified, to the fixed point value which can be extracted from (\ref{3.21})
\be \label{3.51}
N_{Mb;m_1,m_2,m_3}=<N_{,b}\;\chi^M_{m_1}\;\chi^M_{m_2}\; \chi^M_{m_3}>
+
<N\;\chi^M_{m_1}\;[ (\partial_b\chi^M_{m_2})\;\chi^M_{m_3}+\;\chi^M_{m_2}(\partial_b\chi^M_{m_3})]>
\ee
where $\partial_b$ is the continuum derivative. It is obvious that each of the three terms in the 
second line of (\ref{3.50}) corresponds to each of three terms in (\ref{3.51}) in the same order. 
Moreover, the flow equations (\ref{3.39}) are linear in the couplings. Thus convergence can be 
studied for each of three terms separately. Moreover, the second and third term differ only by 
the relabelling of $m_2,m_3$, hence it suffices to consider only the first and second term. 

A technical assumption we will make is that $N$ has {\it compact momentum support}. This will
simplify some of the estimates below. We will comment on later what would need to be done in order to 
lift this restriction. Technically the restriction means that the Fourier coefficients 
of $N$ and thus also of $N_{,b}$ vanish for $n$ outside a compact set in $\mathbb{Z}^3$. 
Thus we find $M_0$ such that all Fourier coefficients vanish when outside $\mathbb{Z}_{M_0}^3$.

\subsubsection{First term}
\label{s3.2.1}

We abbreviate $F(x):=N_{,b}(x)$ and perform the flow for each $b=1,2,3$ separately. The
first iteration gives 
\ba \label{3.52}
F^{(1)}_{M,m} &=&
[(I_{M3M}^\dagger \times I_{M3M}^\dagger \times I_{M3M}^\dagger) \cdot F^{(0)}_{3M,.}](m)
\nonumber\\
&=& \frac{(3M)^6}{(3M)^9}\sum_{m'\in \mathbb{N}_{3M}^9}\; \prod_{s=1}^3\; <\chi^{3M}_{m'_s},\chi^M_{m_s}>
F(x^{3M}_{m'_1})\delta_{m'_2,m'_1}\delta_{m'_3,m'_1}
\nonumber\\
&=& \frac{1}{(3M)^3}\;\sum_{m'\in \mathbb{N}_{3M}^3}\; \prod_{s=1}^3\; <\chi^{3M}_{m'},\chi^M_{m_s}>
F(x^{3M}_{m'})
\nonumber\\
&=& 
\frac{1}{(3M)^3}\;\sum_{n\in \mathbb{Z}_M^9} \;\sum_{m'\in \mathbb{N}_{3M}^3}\; 
[\prod_{s=1}^3\; e^{2\pi i\;n_s[x^{3M}_{m'}-x^M_{m_s}]}\;
F(x^{3M}_{m'})
\nonumber\\
&=& 
\frac{1}{(3M)^3}\;\sum_{n_0\in\mathbb{Z}_{M_0}^3}\;\hat{F}(n_0)\;
\sum_{n\in \mathbb{Z}_M^9} \;
[\prod_{s=1}^3\; e^{-2\pi i\;n_s\;x^M_{m_s}}\;
\;\sum_{m'\in \mathbb{N}_{3M}^3}\; e^{2\pi i x^{3M}_{m'}[n_0+n_1+n_2+n_3]}
\nonumber\\
&=& 
\sum_{n_0\mathbb{Z}_{M_0}^3}\;\hat{F}(n_0)\;
\sum_{n\in \mathbb{Z}_M^9} \;
[\prod_{s=1}^3\; e^{-2\pi i\;n_s\;x^M_{m_s}}]\;\delta_{n_0+n_1+n_2+n_3,0\;({\sf mod}\;3M)}
\ea
where we have introduced the Fourier transform $\hat{F}(n)=<e^{2\pi n\cdot .},F>$ of $F$ and 
implemented the compact support of $\hat{F}$. Note the important modulo operation in the 
Kronecker symbol which results from summation over $m'$ which equals $(3M)^3$ whenever the sum 
of integers displayed is a point in the sublattice of $\mathbb{Z}^3$ whose coordinates are 
integer multiples of $3M$ and otherwise it vanishes.   

The second iteration yields 
\ba \label{3.53}
F^{(2)}_{M,m} &=&
[(I_{M3M}^\dagger \times I_{M3M}^\dagger \times I_{M3M}^\dagger) \cdot F^{(1)}_{3M,.}](m)
\nonumber\\
&=& \frac{1}{(3M)^9}\sum_{m'\in \mathbb{N}_{3M}^9}\; \prod_{s=1}^3\; <\chi^{3M}_{m'_s},\chi^M_{m_s}>
F^{(1)}_{3M,m'}
\nonumber\\
&=& \frac{1}{(3M)^9}\sum_{\hat{n}\in \mathbb{Z}_M^9}\;
\sum_{m'\in \mathbb{N}_{3M}^9}\; [\prod_{s=1}^3\; e^{2\pi i\;\hat{n}_s[x^{3M}_{m'}-x^M_{m_s}]}]\;
F^{(1)}_{3M,m'}
\nonumber\\ 
&=& \frac{1}{(3M)^9}
\sum_{n_0\in\mathbb{Z}_{M_0}^3}\;\hat{F}(n_0)\;
\sum_{n\in \mathbb{Z}_{3M}^9} \;\delta_{n_0+n_1+n_2+n_3,0\;({\sf mod}\;9M)}
\sum_{\hat{n}\in \mathbb{Z}_M^9}\;
\sum_{m'\in \mathbb{N}_{3M}^9}\; [\prod_{s=1}^3\; e^{2\pi i\;(\hat{n}_s[x^{3M}_{m'}-x^M_{m_s}]-n_s x^{3M}_{m'_s}])}]\;
\nonumber\\ 
&=& 
\sum_{n_0\in\mathbb{Z}_{M_0}^3}\;\hat{F}(n_0)\;
\sum_{n\in \mathbb{Z}_{3M}^9} \;\delta_{n_0+n_1+n_2+n_3,0\;({\sf mod}\;9M)}
\sum_{\hat{n}\in \mathbb{Z}_M^9}\;\delta_{n,\hat{n}\;({\sf mod}\;3M)}\;
[\prod_{s=1}^3\; e^{-2\pi i\;\hat{n}_s\; x^M_{m_s}}]
\nonumber\\ 
&=& 
\sum_{n_0\in\mathbb{Z}_{M_0}^3}\;\hat{F}(n_0)\;
\sum_{n\in \mathbb{Z}_M^9} \;\delta_{n_0+n_1+n_2+n_3,0\;({\sf mod}\;9M)}
[\prod_{s=1}^3\; e^{-2\pi i\;n_s\; x^M_{m_s}}]
\ea 
Here we noted that $n=\hat{n}$ modulo $3M$ for $n\in \mathbb{Z}_{3M}^9, \hat{n}\in \mathbb{Z}_M^9$
means $n_s^a=\hat{n}^a_s$ modulo $3M$ for $s,a=1,2,3$ but that $|n^a_s-\hat{n}^a_s|<\frac{3M-1+M-1}{2}<2M<3M$
thus the only solution is $n=\hat{n}$ which was used in the last step. 

Comparing (\ref{3.52}) and (\ref{3.53})
we see that these two expressions differ only by the modulo operation. It follows 
\be \label{3.54}
F^{(r)}_{M,m}=\sum_{n_0\in\mathbb{Z}_{M_0}^3}\;\hat{F}(n_0)\;
\sum_{n\in \mathbb{Z}_M^9} \;\delta_{n_0+n_1+n_2+n_3,0\;({\sf mod}\;3^r\;M)}
[\prod_{s=1}^3\; e^{-2\pi i\;n_s\; x^M_{m_s}}]
\ee
Now $|n^a_0+n^a_1+n^a_2+n_3|\le \frac{M_0+3M-4}{2}$ which is lower than $3^r M$ for 
\be \label{3.55}
r_{M_0,M}:=1+[\frac{\ln(\frac{M_0+3M-4}{2M})}{\ln(3)}]
\ee
where $[.]$ denotes the Gauss bracket. Thus for any $r\ge r_{M_0,M}$ we have 
\be \label{3.56}
F^{(r)}_{M,m}=\sum_{n_0\in\mathbb{Z}_{M_0}^3}\;\hat{F}(n_0)\;
\sum_{n\in \mathbb{Z}_M^9} \;\delta_{n_0+n_1+n_2+n_3,0}
[\prod_{s=1}^3\; e^{-2\pi i\;n_s\; x^M_{m_s}}]
\ee
This is to be compared with the fixed point value 
\ba \label{3.57}
F_{M,m} &=& <F\;\prod_s \chi^M_{m_s}>
=\sum_{n_0\in\mathbb{Z}_{M_0}^3}\;\hat{F}(n_0)\;
\sum_{n\in \mathbb{Z}_M^9} [\prod_{s=1}^3\; e^{-2\pi i\;n_s\; x^M_{m_s}}]
<e^{2\pi\;i[n_0+n_1+n_2+n_3]\cdot .}>
\nonumber\\
&=& \sum_{n_0\in\mathbb{Z}_{M_0}^3}\;\hat{F}(n_0)\;
\sum_{n\in \mathbb{Z}_M^9} \;\delta_{n_0+n_1+n_2+n_3,0}\;
[\prod_{s=1}^3\; e^{-2\pi i\;n_s\; x^M_{m_s}}]
\ea
which equals (\ref{3.56}). 

Thus the convergence of the coupling at fixed resolution $M$ and fixed momentum 
support $M_0$ is uniform in $m\in \mathbb{N}_M^9$. Only finitely many renormalisation 
steps (at most $r_{M_0,M}$) have to be performed before the coupling attains its 
fixed point value. In particular, for resolution $M$ larger than the momentum 
suport we have $r_{M_0,M}=1$, only one step is needed. 

\subsubsection{Second term}
\label{s3.2.2}

The first iteration gives 
\ba \label{3.58}
N^{(1)}_{Mb,m} &=&
[(I_{M3M}^\dagger \times I_{M3M}^\dagger \times I_{M3M}^\dagger) \cdot N^{(0)}_{3Mb;.}](m)
\\
&=& \frac{1}{(3M)^9}\sum_{m'\in \mathbb{N}_{3M}^9}\; [\prod_{s=1}^3\; <\chi^{3M}_{m'_s},\chi^M_{m_s}>]\;
N^{(0)}_{3Mb,m'}
\nonumber\\
&=& -\frac{(3M)^6}{(3M)^9}\sum_{m'\in \mathbb{N}_{3M}^9}\; [\prod_{s=1}^3\; <\chi^{3M}_{m'_s},\chi^M_{m_s}>]\;
N(x^{3M}_{m'_1})\;[\partial_{3Mb} \delta_{m'_1}](m'_2)\;\delta_{m'_1,m'_3}
\nonumber\\
&=& \frac{1}{(3M)^3}\sum_{m'\in \mathbb{N}_{3M}^9}\; [\prod_{2\not=s=1}^3\; <\chi^{3M}_{m'_s},\chi^M_{m_s}>]\;
<[\partial_{3Mb} \chi^{3M}_{.}]_{m'_2},\chi^M_{m_2}>\;
N(x^{3M}_{m'_1})\;\delta_{m'_1,m'_2}\;\delta_{m'_1,m'_3}
\nonumber\\
&=& \frac{1}{(3M)^3}\sum_{m'\in \mathbb{N}_{3M}^3}\; [\prod_{2\not=s=1}^3\; <\chi^{3M}_{m'},\chi^M_{m_s}>]\;
<[\partial_{3Mb} \chi^{3M}_{.}]_{m'},\chi^M_{m_2}>\;
N(x^{3M}_{m'})
\nonumber\\
&=& \frac{1}{(3M)^3}\sum_{n_0\in \mathbb{Z}_{M_0}^3}\;\hat{N}(n_0)\;\sum_{n\in \mathbb{Z}_M^3}\;
(\frac{3M}{2}[e^{-2\pi in^b_2/(3M)}-e^{2\pi i n^b_2/(3M)}])^\ast\;[\prod_{s=1}^3\; e^{-2\pi\; i\; n_s\; x^M_{m_s}}]
\times
\nonumber\\
&& \sum_{m'\in \mathbb{N}_{3M}^3}\; e^{2\pi i x^{3M}_{m'}\;[n_0+n_1+n_2+n_3]}
\nonumber\\
&=& \sum_{n_0\in \mathbb{Z}_{M_0}^3}\;\hat{N}(n_0)\;\sum_{n\in \mathbb{Z}_M^3}\;
(i\; 3M\;\sin(2\pi n^b_2/(3M))\; [\prod_{s=1}^3\; e^{-2\pi\; i\; n_s\; x^M_{m_s}}]\;
\delta_{n_0+n_1+n_2+n_3,0\;({\sf mod}\; 3M)}
\nonumber
\ea
where in the fourth step we summed by parts (no boundary terms due to discrete periodicity) so that the 
discrete derivative acts on the argument $m'_2$ of $\chi^{3M}_{m'_2}$. Then we carried out explicitly 
the discrete derivative on the lattice of resolution $3M$. Note that the discrete derivative acts 
on the label $m'$ of the functions $e^{2\pi i n'(x-x^{3M}{m'})}$ of which $\chi^{3M}_{m'}(x)$ is a 
linear combination.

For the next iteration step it is convenient to introduce the function and parameter
\be \label{3.59}
f(z)=\frac{\sin(z)}{z},\; z^{3M}_n=\frac{2\pi\; n}{3M}
\ee
in terms of which 
\be \label{3.60}
N^{(1)}_{Mb,m}=\sum_{n_0\in \mathbb{Z}_{M_0}^3}\;\hat{N}(n_0)\;\sum_{n\in \mathbb{Z}_M^9}\;
(2\pi i\;n^b_2 f(z^{3M}_{n^b_2}))\;[\prod_{s=1}^3\; e^{-2\pi\; i\; n_s\; x^M_{m_s}}]\;
\delta_{n_0+n_1+n_2+n_3,0\;({\sf mod}\; 3M)}
\ee
By comparing (\ref{3.60}) with (\ref{3.52}) we see that the only difference between 
the two expressions consists in the additional factor that depends on $n^b_2$. Going 
through literally the same steps as in (\ref{3.53}) one therefore quickly convinces oneself
that the next iteration yields 
\be \label{3.61}
N^{(2)}_{Mb,m}=\sum_{n_0\in \mathbb{Z}_{M_0}^3}\;\hat{N}(n_0)\;\sum_{n\in \mathbb{Z}_M^9}\;
(-2\pi i\;n^b_2 f(z^{9M}_{n^b_2}))\;[\prod_{s=1}^3\; e^{-2\pi\; i\; n_s\; x^M_{m_s}}]\;
\delta_{n_0+n_1+n_2+n_3,0\;({\sf mod}\; 9M)}
\ee
and thus 
\be \label{3.62}
N^{(r)}_{Mb,m}=\sum_{n_0\in \mathbb{Z}_{M_0}^3}\;\hat{N}(n_0)\;\sum_{n\in \mathbb{Z}_M^9}\;
(2\pi i\;n^b_2 f(z^{3^r\;M}_{n^b_2}))\;[\prod_{s=1}^3\; e^{-2\pi\; i\; n_s\; x^M_{m_s}}]\;
\delta_{n_0+n_1+n_2+n_3,0\;({\sf mod}\; 3^r M)}
\ee

For $r\ge r_{M_0,M}$ defined in (\ref{3.55}) this simplifies to   
\be \label{3.63}
N^{(r)}_{Mb,m}=\sum_{n_0\in \mathbb{Z}_{M_0}^3}\;\hat{N}(n_0)\;\sum_{n\in \mathbb{Z}_M^9}\;
(2\pi i\;n^b_2 f(z^{3^r\;M}_{n^b_2}))\;[\prod_{s=1}^3\; e^{-2\pi\; i\; n_s\; x^M_{m_s}}]\;
\delta_{n_0+n_1+n_2+n_3,0}
\ee
which is to be compared with the fixed point value 
\be \label{3.64}
N_{Mb,m}=<N\;\;\chi^M_{m_1}\; [\partial_b \chi^M_{m_2}]\; \chi^M_{m_3}>
=
\sum_{n_0\in \mathbb{Z}_{M_0}^3}\;\hat{N}(n_0)\;\sum_{n\in \mathbb{Z}_M^9}\;
(2\pi i\;n^b_2)\;[\prod_{s=1}^3\; e^{-2\pi\; i\; n_s\; x^M_{m_s}}]\;
\delta_{n_0+n_1+n_2+n_3,0}
\ee
It follows
\be \label{3.65}
N^{(r)}_{Mb,m}-N_{Mb,m}=
\sum_{n_0\in \mathbb{Z}_{M_0}^3}\;\hat{N}(n_0)\;\sum_{n\in \mathbb{Z}_M^9}\;
(2\pi i\;n^b_2)\;[f(z^{3^r\;M}_{n^b_2}))-1]\;[\prod_{s=1}^3\; e^{-2\pi\; i\; n_s\; x^M_{m_s}}]\;
\delta_{n_0+n_1+n_2+n_3,0}
\ee
We estimate its modulus 
\be \label{3.66}
|N^{(r)}_{Mb,m}-N_{Mb,m}|\le 2\pi\; 
\sum_{n_0\in \mathbb{Z}_{M_0}^3}\;|\hat{N}(n_0)|\;\sum_{n\in \mathbb{Z}_M^9}\;
|n^b_2|\;|f(z^{3^r\;M}_{n^b_2}))-1|\;
\delta_{n_0+n_1+n_2+n_3,0}
\ee
which is independent of $m\in \mathbb{N}_M^9$. We now use the fact that $f(z)=f(-z)$ and 
\be \label{3.67} 
|f(z)-1|\le z
\ee
for all $z\ge 0$ which can be shown by proving that $g_\pm(z):=z^2\pm (z-\sin(z))\ge 0$
(take first derivative of $g_+$ and second derivative of $g_-$ to prove strict monotonicity
of $g_\pm$).  Then (\ref{3.66}) can be further estimated by 
\be \label{3.68}
|N^{(r)}_{Mb,m}-N_{Mb,m}|\le 2\pi\;||\hat{N}||\; 
\sum_{n_0\in \mathbb{Z}_{M_0}^3}\;\sum_{n\in \mathbb{Z}_M^9}\;
|n^b_2|\;|z^{3^r\;M}_{n^b_2}|\;
\delta_{n_0+n_1+n_2+n_3,0}
\ee
with $||\hat{N}||={\sf sup}_{n_0\in \mathbb{Z}_{M_0}^3}\; |\hat{N}(n_0)|$. In carrying 
out the Kronecker in (\ref{3.68}) we cannot simply solve for say $n_1=-(n_0+n_2+n_3)$ because 
$n_1$ is subject to the constraint $n_1\in \mathbb{Z}_M^3$. However, ignoring that constraint 
simply gives more positive terms so that certainly 
\ba \label{3.69}
&& |N^{(r)}_{Mb,m}-N_{Mb,m}| \le  2\pi\;||\hat{N}||\; 
\sum_{n_0\in \mathbb{Z}_{M_0}^3}\;\sum_{n_2,n_3\in \mathbb{Z}_M^3}\;
|n^b_2|\;|z^{3^r\;M_{n^b_2}}|
\nonumber\\
&\le & 2\pi\;||\hat{N}||\; M_0^3\; M^5\;
\sum_{n_2^b\in \mathbb{Z}_M}\;
\frac{2\pi [n^b_2]^2}{3^r M}
\nonumber\\
&\le & 3^{-r}\; \frac{[2\pi]^2}{12}\;||\hat{N}||\; M_0^3\; M^4\; (M+1)^3\;
\ea   
where we used 
\be \label{3.70}
\sum_{|n|\le (M-1)/2} n^2=2\;\sum_{n=1}^{(M-1)/2}\; n^2\le 
2\;\sum_{n=1}^{(M-1)/2}\; \int_n^{n+1}dx\; x^2
\le 2\int_0^{(M+1)/2} \; dx\; x^2=\frac{(M+1)^3}{12}
\ee
Given $\epsilon>0$, we pick 
\be \label{3.71}
r_{\epsilon,||\hat{N}||,M_0,M}:=1+[\frac{\ln(\pi^2\; ||\hat{N}||\; M_0^3\; M^4\; (M+1)^3/3)}{\ln(3)}]
\ee
then $|N^{(r)}_{M,m}-N_{M,m}|\le \epsilon$ for $r\ge r_{\epsilon,||\hat{N}||,M_0,M}$. Note 
that the convergence is again uniform in $m\in \mathbb{N}_M^9$. The convergence is in fact 
exponentially fast at given $N,M,M_0$. 

\subsubsection{Lifting the compact momentum support}
\label{s3.2.3}

If one drops the restriction of compact momentum support, one has to impose weaker 
decay properties on the Fourier transform $\hat{N}$. The estimates performed above
relied on the fact that at fixed $M$ eventually $n_0+n_1+n_2+n_3$ cannot exceed $3^r M$ as $r$ grows  
when $|n_0^b|\le (M_0-1)/2$ is bounded. If such $M_0$ does not exist, but $\hat{N}(n_0), \;n_0^b\; \hat{N}(n_0)$
drop sufficiently fast as $n_0\to\infty$ then the additional solutions of $n_0+n_1+n_2+n_3=0$ modulo 
$3^r M$ at fixed $M$ as compared to $n_0+n_1+n_2+n_3=0$ necessarily involve large $n_0$ of the order 
of $3^r M$ as $r\to \infty$ because $|n^b_1+^b_2+n^b_3|<3M/2$. Those additional contributions then 
are small if one imposes e.g. rapid decay on $\hat{N}$ (i.e. 
$||\hat{N}||_{p}=\sup_{n\in \mathbb{Z}^3}\; |[n^1]^{p_1}\;[n^2]^{p_2};[n^3]^{p_3} \hat{N}(n)|<\infty$
for all $p_1,p_2,p_3\in \mathbb{N}_0$). The other estimate that needs to be reconsidered is 
that $\sum_{n_0} |\hat{N}(n_0)|$ can no longer be estimated by $M_0^3 ||\hat{N}||$, however, if 
$\hat{N}(n_0)$ has rapid decay then this sum certainly exists. We leave the details to the 
interested reader.

\subsection{Discussion}
\label{s3.3}

We end this section by noting a few observations.
\begin{itemize}
\item[1.]
As we have shown, different initial data such as the ones suggested in (\ref{3.28}) are not a fixed point of (\ref{3.39}). In 
\cite{RZ-T} similar flow equations as for $N_{M,b}$ appear which there however do not involve spatial derivatives and the analog
of the lapse functions is just the constant function equal to unity. There the correct fixed point value of the 
coupling is the integral over a product of $\chi^M_m$ functions which is close to a ``naive'' starting point 
given by a product of Kronecker $\delta$'s. One can show in that case that the flow with naive initial data 
reaches the correct fixed point after finitely many steps where the step number 
depends only on the polynomial degree of the interaction polynomial. In the present situation, the situation is more complicated
due to the appearance of the spatial derivative and the fact that the functions $N, u^a$ are not constant. 
The naive starting point to locate $N,u^a$ at the points $x^M_m=m/M$ and to use one of the typical lattice derivatives 
such as forward, backward or antisymmetric derivative with summation kernel
\be \label{3.42a}
\partial^s_{M,a}(m_1,m_2)=\frac{M(1+|s|)}{2}[\delta_{m_1+[1+\frac{s-|s|}{2}]\delta_a,m_2}
-[1-\frac{s+|s|}{2}]\delta_{m_1+\delta_a,m_2}]
\ee
with $s=1,-1,0$ prevent the flow from converging after finitely many steps only. 
By contrast, using the natural lattice derivative $\partial_{M,a}=I_M^\dagger \partial_a I_M$ that enjoys the intertwining identity 
$\partial_{3M,a} I_{M3M}=I_{M3M} \partial_{M,a}$ accelerates convergence.
In order to show this, it is crucial that $D[u], H[N]$ are polynomials as otherwise 
the integral over $x$ cannot be performed in closed form. Otherwise we obtain a non-polynomial dependence
on $I_{M3M}$ and the basic mechanism (\ref{3.41}) cannot be used in order to 
manipulate the flow equations corresponding to (\ref{3.39}).
\item[2.]
The non-triviality 
of the flow equations in the discrete ``real space'' framework is due to the non-triviality of the embedding 
$I_{M3M}:\;l_M\to l_{3M}$. By contrast, in the projected 
framework the embedding $L_M\to L_{3M}$ is trivial and thus the flow equations are trivial. 
It is however a feature specific to the Narnhofer-Thirring representation that our discretisation 
prescription to replace $A,E$ by $A_M=P_M\cdot A,E_M=P_M\cdot E$ 
and leave $\partial_a$ untouched in $H[N]$ to obtain $H_M[N]$ as our initial datum in fact coincides with the 
constraint blocked from the continuum: As far as $E$ is concerned, this is clear from the fact that 
$W_M[F_M]\Omega_M$ are eigenstates of both $E,E_M$ with eigenvalues $F_M$ and would not hold in other 
representations in which $E$ is not diagonal. As far as $A$ is concerned this is true because in decomposing 
$A=A_M+A_M^\perp$ and regularising both contributions in terms of 
Weyl elements the contribution $A_M^\perp$ drops from matrix elements between vectors of the form $W_M[F_M]\Omega_M$
because of the specific feature of the Narnhofer-Thirring representation that such vectors form an orthonormal 
basis which fails to hold in more regular representations. This works for any density weight and holds as long as $P_M$ is a 
smooth projection $P_M:\; L\to L_M$ such that $\partial_a L_M\subset L_M$ and does not depend on the 
specific form of $P_M$.
\item[3.]
Clearly, by rewriting $H_M[N]$, defined as $H[N]$ with the substitutions $(A,E)\to (A_M=P_M\cdot A,E_M=P_M\cdot E)$ and then 
rewriting $(A_M,E_M) = (I_M\cdot a_M, E_M=I_M\cdot e_M)$ automatically replaces $\partial_a$ by 
$\partial_{M,a}=I_M^\dagger\partial_a I_M$ as $\partial_a F_M=P_M \cdot\partial_a F_M =I_M \cdot\partial_{M,a} f_M,\;
F_M=I_M\cdot f_M$ whenever $\partial_a L_M\subset L_M$. This is not true for instance when we use the projection 
based on the Haar kernel. It follows that the thus reformulated $H_M[N]$ written in terms of $a_M, e_M, \partial_M$ is 
also a fixed point of the flow equations for any density weight $w$ in the discretised setting. 
This is true, however, the flow equations 
rewritten in terms of the couplings now take a much more complicated form than for the case $w-2=4k,\; k\in \mathbb{N}_0$.
This is because in the discretised setting we want to cast the flow equations in terms of the couplings
depending only on the discretised labels $m\in \mathbb{N}_M^3$. In order to achieve this, we must carry 
out the $x$ integral appearing in $H_M[N]$ and for $w-2\not=4k$ this can no longer be done non-perturbatively
because of the non-polynomial dependence of $H_M[N]$ in the functions $I_M$. At best, one 
can hope to aim for a perturbative treatment.
\item[4.]
To see 
this explicitly, recall that for general density weight 
$w$ the Hamiltonian constraint blocked from the continuum takes the form
\be \label{3.42}
H_M[N]=\int\; d^3x\; N(x)\; A^j_{M,a}(x)\;\epsilon_{jkl}\;[|\det(E_M(x))|^{(w-2)/2}\; E^a_{M,k}(x) E^b_{M,l}(x)]_{,b}
\ee
In order to write this explicitly in terms of the $a_m=I_M^\dagger\cdot A_M, e_M=I_M^\dagger \cdot E_M$
we simply substitute $A_M=I_M\cdot a_m, E_M=I_M\cdot e_M$. In the discrete picture one would 
like to integrate out 
the $x$ dependence in order to obtain a flow equation in terms couplings depending on 
discretised points $m\in \mathbb{N}_M^3$. If $w-2=4k,\; k=0,1,2,..$ is a non-negative multiple of 4 we obtain 
a flow equation in terms of a couplings $N_{M,b}$ depending on $3(1+2k)$ points and above 
considerations for $k=0$ go through with mild complications. 
However, for any other values of $w$, the situation changes drastically. 
The coefficient of $N$ of the integrand of (\ref{3.42}) now does not belong of any finite dimensional 
subspace $L_{M'},\; M'<\infty$ while for $w-2=4k$ it belongs to $L_{3(1+2k)M}$. As an illustrative example,
consider the case that $w=-2$. Then we are interested in integrals of the form 
\be \label{3.43}
<\frac{\prod_{I=1}^3 \chi^M_{m_I}}{\sum_{m_4,..,m_9} q_{m_4,..,m_9}\; \prod_{J=4}^9\; \chi^M_{m_j}}>
\ee
with $m_1,..,m_9\in \mathbb{N}_M^3$ and $q_{m_4,..,m_9}$ are homogeneous polynomials of the 
$e^a_{M,j}(m_J)$ of order six. Now $\chi^M_m(x)=\sum_{|n|\le (M-1)/2} e^{2\pi i n (x-m/M)}$. 
We write $\chi^M_m=1+\hat{\chi}^M_m$ isolating the homogeneous mode, define 
$q=\sum_{m_4,..,m_9} q_{m_4..m_9}$ and expand 
\be \label{3.44}
q^{-1}\;[1+\sum_{m_4,..,m_9}\frac{q_{m_4,..,m_9}}{q}\;[-1+\prod_{J=4}^9\;(1+\hat{\chi}^M_{m_j})]^{-1}
\ee
in a geometric series. Leaving aside convergence issues of such an expansion it is apparent that 
the resulting integral now depends on an infinite set of couplings labelled by arbitrarily large number of 
discretised points and thus displays a tremendous amount on non-locality, although integrals of the 
form $<\prod_{I=1}^N \chi^M_{m_I}>$ are strongly peaked at $m_1=..=m_N$ whenever the 
$\chi^M_m$ are quasi-local which is true for the Dirichlet kernel.
\item[5.]
This infinite set of couplings take definite, computable values 
as they are blocked from the continuum, the theory is predictive. 
However, one could have used instead of (\ref{3.42}) for $w=-2$ the naive expression
\be \label{3.44a}
H^{{\sf loc}}_M[N]=M^{-3}\sum_{m\in \mathbb{N}_M^3}\; N(x^M_m)\; a^j_{m,a}(x)\;\epsilon_{jkl}\;
(\partial_{M,b}[\det(e_M)]^{-2}\; e^a_{M,k} e^b_{M,l}])(m)
\ee 
with say the forward lattice derivative $\partial_M$. This expression is no fixed point of the resulting flow equations as 
it involves couplings depending on only two points (it would be only one were it not for the derivative).
It is rather likely that an infinite number of iterations of the resulting flow equations would need 
to be performed in order to show that (\ref{3.44}) flows into (\ref{3.42}).
\item[6.]
This issue is of considerable interest because of the following reason: \\
A non-polynomial Hamiltonian constraint
is in fact strongly motivated by the full SU(2) theory where density weight $w=1$ of $H[N]$ is preferred 
\cite{q}. However, for the SU(2) theory the exact continuum theory is unknown. Still, rather local, discretised expressions 
similar to (\ref{3.44}) have been used as starting points to define the quantum dynamics \cite{q}, albeit 
with different choices of discretisations $I_M$ to construct $a_M=I_M^\dagger \cdot A, e_M=I_M^\dagger \cdot E$
(instead of smearing in all three directions one smears in only one or two directions respectively to 
obtain holonomies and exponentiated fluxes as Weyl elements). 
The above considerations suggest that such a rather local starting point is much too restrictive 
in order to find the correct continuum theory. It is conceivable
that the correct SU(2) continuum theory therefore takes a rather complicated non-local form when blocked 
from the continuum to resolution $M$ and written in the discretised variables $a_M, e_M$. Vice versa, it 
could take 
a much simpler form when written in the projected variables $A_M = P_M\cdot A, E_M=P_M\cdot E$ 
because it is not necessary to carry out the $x$ integral. 
\end{itemize}

\section{The algebroid flow}
\label{s4}

The renormalisation flow for the 
algebroid solution benefits from previous works \cite{RZ-T} or \cite{i}. In the first subsection we discuss the implementation 
via \cite{i}. This uses 1. a flat background and 2. a trivial covariance. This framework shows that not only the quadratic forms 
do flow to their correct limit but also that the constraint algebra closes without anomalies. Here we work in the projection 
formalism. In the second subsection we consider 
the implementation via \cite{RZ-T} which only uses the flat background but allows for more general covariances. Checking whether the 
algebra still closes in this case is left for future investigation. Here we work in the discretisation formalism.   

The general setting is as follows: We start by fixing an arbitrary background metric $g$ with Euclidean signature on $\sigma$. 
The scalar one-particle Hilbert 
space is given by \(L_2:=(\sigma,\sqrt{\det(g)}\;d^3x)\). 
We consider the Laplacian $\Delta=g^{ab}\nabla_a\nabla_b$, where $\nabla_a$ is the Levi-Civita covariant derivative
of $g$. We consider a strictly positive, hence invertible, function  $\kappa$ of the smooth, self-adjoint
operator $\Delta$. Then
the annihilation operator is given by
\be \label{4.1}
B_{aj}:=\frac{1}{\sqrt{2}}(\delta_{jk}\kappa A_a^k-i\kappa^{-1}\omega^{-1} g_{ab} E^b_j),
\ee
where $\omega = \sqrt{\det(g)}$ denotes the volume form of $g$. For the first implementation we specialise to a flat background metric 
$g_{ab} = \delta_{ab}$ and trivial covariance $\kappa=1$ while in the second implementation we allow for a general
translation invariant covariance $\kappa$ while still keeping a flat background. 
Then we are precisely within the frameworks of \cite{i,RZ-T} respectively 
and can directly apply the machinery developed therein
for the quantum constraints $C[r],\; D[u],\; H[N]$, which coincide with their classical expressions 
(with density weight $w=2$ for $H$) except that one has to rewrite them in terms of 
$B_{aj},\; [B_{aj}]^\ast$ and normal order. 

\subsection{Trivial covariance in terms of projections}
\label{s4.1}

If we consider $\kappa=1$, with  $C_M[r], \; D_M[u], \; H_M[N]$ 
being the same as $C[r],\; D[u],\; H[N]$ but with $B_{aj},\; [B_{aj}]^\ast$ replaced by their projections 
$P_M\cdot B_{aj},\; [P_M B_{aj}]^\ast$ where 
$P_M$ is the Dirichlet kernel of appendix \ref{sa}, then we are in a particular incarnation 
of \cite{i} in the sense that in \cite{i} the annihilation and creation operators were 
expanded in terms of an arbitrary smooth real valued ONB of the 1-particle Hilbert subspace $L_M$ of $L$  
(here $L=L_2([0,1)^3,d^3x)$) while here we use the specific ONB $b^M_m=\prod_{a=1}^3\; [\chi^M_{m_a}(x^a)/\sqrt{M}],\;
m_a\in \mathbb{N}_M$ and decompose $P_M=\sum_{m\in \mathbb{N}_M}^3 \; b_m\; <b_m,.>_L$. 
We may therefore use the result of \cite{i} which shows that there 
exists a limiting pattern in the sense of section \ref{s2.3} such that the quadratic form 
commutators cut-off at finite $M$ converge in the weak Fock Hilbert space topology to 
normal ordered quadratic forms that precisely coincide with the normal ordered corresponding 
Poisson brackets times $i$. It follows that the algebroid flow in terms of projections 
has the solution \cite{i} as fixed point. This fixed is point is reached already at the zeroth step,
that is, the constraints with $A,E$ replaced by $A_M=P_M \cdot A, E_M=P_M\cdot E$ and normal ordered result 
in the same quadratic form as the one that results by blocking from the continuum.
To see this one decomposes the continuum operator in terms $B, B^\dagger$. Then sandwiching the continuum operator 
between states of he form $W[P_\cdot F]\Omega$ yields a polynomial in functions of the form 
$P_M F$. But since since the annihilation operator $B_M$ for $\kappa=1$ built from $A_M,E_M$ 
is given by (\ref{4.1}) with the substitution $(A,E)\to (A_M,E_M)$ we simply have 
$B_M=P_M\cdot B$. This means that normal ordering at resolution $M$ yields the same normal ordering
as in the continuum. Finally since $P_M^2=P_M$ is a projection, the resulting expression when 
sandwiched between states of the form $W[P_M F]\Omega$ produces the same polynomial in the 
functions $P_M\cdot F$ as the continuum operator. Since we leave the smearing functions 
$r,u,N$ untouched the resulting quadratic forms trivially coincide.
Finally, 
we may copy the limiting pattern established in \cite{i} to perform the weak limit of finite resolution quadratic form
commutators to conclude that there is no anomaly in the constraint algebra. 

\subsection{Translation invariant covariance in terms of discretisations}
\label{s4.2}

Consider a flat background and $\kappa$ a general 
strictly positive, operator valued function of the corresponding translation invariant, self-adjoint Laplacian $\Delta$. 
We discretise the constraints as in section \ref{s3.2} with density weight $w=2$ of the Hamiltonian 
constraint (i.e. $H=H_2:=2\epsilon^{jkl}\partial_{[b}A^j_{c]}E^b_kE^c_l$). 
The idea is to exploit the observation made in \cite{RZ-T} that whenever one has a quadratic 
form $H$ on a Fock space ${\cal H}$ which is constructed using annihilation operators 
of the form $a=2^{-1/2}[\kappa\cdot \phi-i\kappa^{-1}\cdot \pi]$ where $\phi,\pi$ are
canonically conjugate variables and $\kappa$ is a translation invariant, strictly positive operator
on the one particle Hilbert space (i.e. it commutes with all spatial derivatives), 
then, the renormalisation flow of the theory compactified on the $D$-torus ($D$ being the spatial dimension) and using the Dirichlet 
kernel and the discretisation scheme described in \cref{sa} admits the continuum theory $(\mathcal{H}, H)$ as a fixed point.
Furthermore, merely substituting $E_M=I_M e_M,\; A_M= I_M a_M$ 
into the expressions of the previous subsection does not produce anything new except that we deal with a non-trivial 
kernel $\kappa$. The resulting flow still converges already at the zeroth step due to the identity $\partial_a I_M=I_M \partial_{M,a}$ and the 
fact that $\Delta_M=\delta^{ab} \partial_{M,a}\partial_{M,b}$.
However, using the localised initial data of the flow such as (\ref{3.50}) displays  
a less rapidly converging flow that gets fixed at the correct, quasi-local fixed point. This will be established in the following.\\  
\\
Since the calculations are rather similar to those of the proceeding section and to \cite{RZ-T}, we can be brief and refer 
the reader to \cite{RZ-T} for more details. We 
define the discretised annihilation operator by 
\be \label{4.2}
b_{M,aj}=\frac{1}{\sqrt{2}}(\delta_{jk}\kappa_M a_{Ma}^k-i\kappa_M^{-1}\delta_{ab} e^b_j),
\ee
where for some strictly positive function $\kappa$
\be \label{4.3}
\kappa_M=\kappa(\Delta_M),\Delta_M=\delta^{ab}\partial_{M,a} \partial_{M,b},\; 
\partial_{M,a}=I_M^\dagger\cdot \partial_a \cdot I_M
\ee
As shown in the appendix $\partial_{3M,a}\cdot I_{M3M}=I_{M3M}\cdot \partial_{M,a}$ hence by the 
spectral theorem 
\be \label{4.4}
I_{M3M}\cdot \kappa_M^{\pm 1}=\kappa_{3M}^{\pm 1} \cdot I_{M3M} 
\ee
as all derivatives mutually commute. 
The idea is now to expand the concrete expression (\ref{3.50}) in terms of annihilation 
and creation operators and then to normal order. Thus we write 
\be \label{4.5}
a_{Ma}^j=\frac{1}{\sqrt{2}} \kappa_M^{-1}\cdot \delta^{jk}\;[b_{M,ak}+b_{M,ak}^\dagger],\;
e^a_{Mj}=i\frac{1}{\sqrt{2}} \kappa_M \cdot \delta^{ac}\;[b_{M,cj}-b_{M,cj}^\dagger]
\ee
plug this into (\ref{3.50}) and then normal order.

Since the Fock Hilbert space is defined in terms of the covariance $\omega_M$ of its Gaussian measure 
which in turn is a function of $\kappa_M$, it follows 
$I_{M3M}^\dagger \cdot \omega_{3M} \cdot I_{M3M} =\omega_M$ where $I_{M3M}^\dagger \cdot I_{M3M}=1_{l_M}$ 
was used. Hence the Fock representations at reesolution $M$ are expectedly already at their fixed point. Next we 
have with Weyl elements $w_M[f_M]=\exp(-i<f_M,a_M>_{l_M^9})$ 
\ba \label{4.6}
&& b_{3M, aj}(m')\; w_{3M}[I_{M3M}\cdot f_M]\;\Omega_{3M}
\nonumber\\
&=& w_{3M}[I_{M3M}\cdot f_M]\; 
(b_{3M,aj}(m')+i[<I_{M3M}\cdot f_M,a_{3M}>_{l_{3M}^9},b_{3M,aj}(m')])\;\Omega_{3M}   
\nonumber\\
&=& \frac{i}{\sqrt{2}}\;w_{3M}[I_{M3M}\cdot f_M]\; 
[<\kappa_{3M}^{-1}\cdot I_{M3M}\cdot f_M,(b_{3M}+b_{3M}^\dagger)>_{l_{3M}^9},b_{3M,aj}(m')]\;\Omega_{3M}   
\nonumber\\
&=& -\frac{i}{\sqrt{2}}\;
[I_{M3M}\cdot\kappa_M^{-1} \cdot\;f_M]_{aj}(m')\;
w_{3M}[I_{M3M}\cdot f_M]\; 
\Omega_{3M}   
\ea
Accordingly, when sandwiching the normal ordered constraint operator consisting of 
polynomials (of order two and three respectively for the 
spatial diffeomorphism and Hamiltonian constraint respectively) 
in\\
$\kappa_{3M}^{\pm 1}\cdot b_{3M,aj},\;\kappa_{3M}^{\pm 1}\cdot b_{3M,aj}^\dagger$ between the states 
$w_{3M}[I_{M3m}\cdot f_M]\Omega_{3M}$, the result of the computation is the same as sandwiching the same
constraint operator consisting of the same polynomial in $\kappa_{M}^{\pm 1}\cdot b_{M,aj},\;\kappa_{M}^{\pm 1}\cdot b_{M,aj}^\dagger$
between the states $w_{M}[f_M]\Omega_{M}$ except that one has to map the resulting factor functions $\kappa_M^n \cdot f_M, n\in \mathbb{Z}$ 
with 
$I_{M3M}$. Each of these factor functions is subject to an $l_M^9$ inner product with respect to one of the entries 
of the coupling function. E.g. $N^{(r)}_{3M c,m'_1, m'_2,m'_3}\delta^{ji}\epsilon_{ikl}\delta^a_b$ is multiplied 
by $a^j_{3M,a}(m'_1)\; e^c_{Mk}(m'_2)\; e^b_{Ml}(m'_3)$, then all indices are contracted and the sum is carried 
out over $\mathbb{N}_{3M}^9$ with weight $(3M)^{-9}$. This is precisely the same as an inner product in 
$l_{3M}^9\otimes l_{3M}^9\otimes l_{3M}^9$. Then taking the adjoint of the occurring $I_{M3M}$ operations we 
can let them act on the coupling rather than the functions $f_M$. Accordingly we end up with exactly the same flow equations 
(\ref{3.39}) as in the groupoid solution case and the fixed point analysis is literally the same.

\section{Conclusions and outlook}
\label{s7}

In the present work we applied the version of Hamiltonian renormalisation developed throughout 
this series of works to U(1)$^3$ General Relativity which is a 3+1 dimensional self-interacting 
QFT modelling Euclidian signature vacuum GR in 3+1 dimensions. The operator constraint quantisations 
of this theory in the continuum were carried out in \cite{g,i} in Narnhofer-Thirring and Fock representations 
respectively. We applied the Hamiltonian renormalisation scheme using initial finite resolution families
of constraints which are adapted to these representations. In particular, we considered finite 
resolution representations of the same type and quantised the coarse grained groupoid and algebroid respectively
at finite resolution just as the continuum constraints. We found that a fixed point exists and 
coincides with the constraints blocked from the continuum. 

The groupoid quantisation 
constructs bounded exponentiated constraint {\it operators} and we could define the exponentiated Hamiltonian constraint for any 
density weight provided the Weyl operators $W[F]$ are smeared with non-degenerate smearing functions 
(i.e. $\det(F)\not=0$) while the algebroid quantisation constructs {\it quadratic forms} and is restricted to density 
weights $w$ such that the Hamiltonian constraint is a polynomial. In both cases the coarse graining was with respect to the Dirichlet kernel  
which projects the operators with respect to their spatial dependence equally for all three directions. The projection 
is equivalent to a smearing or mollification of $A,E$ in all three directions. This is sufficient 
to obtain (exponentiated) operators in the groupoid case because in U(1)$^3$ theory the constraints depend at most 
linearly on momentum (in this case $A$) and thus the constraints preserve the Abelian subalgebra of the Poisson algebra 
of functions on the phase space that depend on the configuration coordinates (in this case $E$) only. For the actual
non-Abelian theory which depends quadratically on $A$ this polarisation is no longer preserved. 
The analysis of \cite{q} shows that a more singular coarse graining  
has to be performed if one wants to obtain (exponentiated) operators and that one should pick 
the natural density weight $w=1$: Specifically, $A_a^j$ should be smeared  
only with respect to the $a-$direction while $E^a_j$ should be smeared only with respect to $b,c-$directions with 
$\epsilon_{abc}=1$. This is equivalent to a coarse graining kernel  which for $A_a^j$ consists of a product 
of a Dirichlet kernel for the $a-$direction
and two $\delta-$kernels for $b,c-$directions and for $E^a_j$ it consists of a product a $\delta$ kernel for the $a-$direction
and two Dirichlet kernels for $b,c-$directions respectively. The corresponding renormalisation is much more involved, in particular 
due to the non-polynomial nature of the constraints and the corresponding difficulty to maintain the non-degeneracy 
condition and therefore will be reserved for a future publication. 

On the other hand, it was clear that the algebroid renormalisation of full Euclidian signature GR at $w=2$ using Dirichlet kernels for all 
directions delivers as fixed point solution the solution given in \cite{i} if the Fock representation uses a flat background and a 
trivial
covariance $\kappa$. This is  because all that was used in \cite{i} was  
that one works in Fock representations of polynomial quadratic forms and general orthonormal bases as cut-off smearing 
functions and that is sufficient in order that constraints blocked 
from the continuum are reached as a fixed point of the flow, as shown in \cite{RZ-T} for the example of $P(\Phi)_2$. We  
extended this result explicitly to the case of a flat background and a non-trivial covariance in section \ref{s4.2}. What has not been 
checked yet for non-trivial covariance is whether also the constraint algebra remains free of anomalies (in \cite{i} this was verified 
only for trivial covariance). We hope to fill this gap in a future publication.   

One of the observations made in the present contribution is that renormalisation in Narnhofer-Thirring 
representations has to cope with the discontinuity of the representation in the sense that usual 
renormalisation prescriptions relying on (weak) continuity of matrix elements have to be properly 
reformulated using the inherent discrete topology of the representation. We have shown how this 
can be accomplished.

Finally, note that the algebra of the coarse grained constraints blocked from the continuum do not close at 
any finite resolution $M$ even for trivial covariance in the case of the algebroid solution. 
This is neither expected nor a contradiction as pointed out in \cite{TZ} and simply 
a consequence of the presence of the projections $P_M$. The corresponding anomalies vanish as $M\to \infty$ in 
the weak operator topology \cite{TZ}. The continuum constraints at infinite resolution 
do close in the sense of \cite{g,i}.
\\ 
\\       
{\bf Acknowledgements}\\
M.R.-Z. thanks Jonas Neuser for useful discussions on the topic and acknowledges the financial support provided by the Deutsche Akademische Austauschdienst e. V.\\

\begin{appendix}

\section{Renormalisation tools}
\label{sa}

More details on this section, in particular the relation to wavelet theory \cite{m},  can be found in \cite{l}.\\
\\
We work on spacetimes diffeomorphic to $\mathbb{R}\times\sigma$. In a first step the 
spatial D-manifold $\sigma$ is compactified to $T^D$. Therefore, all constructions 
that follow have to be done direction wise for each copy of $S^1$. On $X:=S^1$, understood 
as $[0,1)$ with endpoints identified, we consider 
the Hilbert space $L=L_2([0,1),\;dx)$ with orthonormal basis 
\be \label{a.1}
e_n(x):= e^{2\pi\;i\;n\;x},\; n\in\mathbb{Z}
\ee
with respect to the inner product 
\be \label{a.2}
<F,G>_L:=\int_0^1\;dx\; \overline{F(x)}\; G(x)
\ee 
Let $\mathbb{O}\subset\mathbb{N}$ be the set of positive odd integers. We equip $\mathbb{O}$ 
with a partial order, namely 
\be \label{a.3}
M<M' \;\;\Leftrightarrow\;\; \frac{M'}{M}\in \mathbb{N}
\ee
Note that this is not a linear order, i.e. not all elements of $\mathbb{O}$ are in relation,
but $\mathbb{O}$ is directed, that is, for each $M,M'\in \mathbb{O}$ we find 
$M^{\prime\prime}\in \mathbb{O}$ such that $M,M'<M^{\prime\prime}$ e.g. $M^{\prime\prime}=M M'$.  
For each $M\in \mathbb{O}$, called a resolution scale, we introduce the subsets 
$\mathbb{N}_M\subset \mathbb{N}_0,\;
\mathbb{Z}_M\subset \mathbb{Z},
X_M\subset X$ of respective cardinality $M$ defined by 
\be \label{a.4}
\mathbb{N}_M=\{0,1,..,M-1\},\; 
\mathbb{Z}_M=\{-\frac{M-1}{2}, -\frac{M-1}{2}+1,..,\frac{M-1}{2}\},\;
X_M=\{x^M_m:=\frac{m}{M},\; m\in \mathbb{N}_M\}        
\ee
It is easy to check that we have the lattice relation 
\be \label{a.5}
X_M\subset X_{M'} \;\;\Leftrightarrow\;\; M<M'
\ee
The subspace $L_M\subset L$ is defined by 
\be \label{a.6}
L_M:={\sf span}(\{e_n,\; n\in \mathbb{Z}_M\})
\ee
On $L_M$ we use the same inner product as on $L$, hence the $e_n,\; n\in \mathbb{Z}_M$ 
provide an ONB for $L_M$. An alternative basis for $L_M$ is defined by the functions
\be \label{a.7}
\chi^M_m(x):=\sum_{n\in \mathbb{Z}_M}\; e_n(x-x^M_m)
\ee
The motivation to introduce these functions is that in contrast to the plane waves 
$e_n$ they are 1. spatially concentrated at $x=x^M_m$ and 2. real valued. This 
makes them useful for renormalisation purposes. In addition, in contrast to 
characteristic functions which have better spatial location properties, they 
are smooth. This is a crucial feature because quantum field theory involves 
products of derivatives of the fields and derivatives of characteristic functions 
yield $\delta$ distributions. More in general, renormalisation tools must make 
a compromise between localisation and smoothness.      

The functions $\chi^M_m$ are still orthogonal but not orthonormal
\be \label{a.8}
<\chi^M_m,\;\chi^M_{\hat{m}}>_{L_M}=M\; \delta_{m,\hat{m}}
\ee
We choose not to normalise them in order to minimise the notational clutter in what 
follows. Let $l_M$ be the space of square summable sequences $f_M=(f_{M,m})_{m\in \mathbb{N}_M}$
with $M$ members and 
inner product given by 
\be \label{a.9}
<f_M,\;g_M>_{l_M}:=\frac{1}{M}\sum_{m\in\mathbb{N}_M}\; \overline{f_{M,m}}\; g_{M,m}  
\ee
If we interpret $f_{M,m}=F(x^M_m)$ then (\ref{a.9}) is a lattice approximant of 
$<F,G>_L$. We define 
\be \label{a.10}
I_M:\; l_M\to L_M;\; (I_M\cdot f_M)(x):=
<\chi^M_{\cdot}(x), f_M>_{l_M}=\frac{1}{M}\;\sum_{m\in \mathbb{N}_M}\; f_{M,m}\; \chi^M_m(x)
\ee
Its adjoint is defined by the requirement that 
\be \label{a.11}
<I_M^\ast\cdot F_M,\; g_M>_{l_M}= <F_M,\; I_M\cdot g_M>_{L_M}
\ee
which demonstrates 
\be \label{a.12}
I_M^\ast:\; L_M\to l_M;\; (I_M^\ast\cdot F_M)_m=<\chi^M_m, F_M>_{L_M}
\ee
One easily checks, using (\ref{a.8}) that 
\be \label{a.13}
I_M^\ast \cdot I_M=1_{l_M}, \;\;<I_M .,I_M .>_{L_M}=<.,.>_{l_M}
\ee
which shows that $L_M, l_M$ are in 1-1 correspondence and that $I_M$ is an isometry.
Likewise 
\be \label{a.13a}
P_M:=I_M \cdot I_M^\ast=1_{L_M}
\ee
We can consider $I_M$ also as a map $I_M:\; l_M\to L$ with image $L_M$ and then 
$<I_M^\ast ., .>_{l_M}=<.,I_M>_L$ shows that $I_M^\ast: L\to l_M$ is given by 
the same formula (\ref{a.12}) with $F\in L$ but now $P_M: L\to L_M$ is an orthogonal 
projection 
\be \label{a.14}
P_M\cdot P_M=P_M,\; P_M^\ast=P_M
\ee
We have explicitly 
\be \label{a.15}
(P_M\cdot F)(x)=\int_0^1\; dy\; P_M(x,y)\; F(y),\;
P_M(x,y)=\sum_{n\in \mathbb{Z}_M}\; e_n(x-y)
\ee
i.e. $P_M(x,y)$ is the M-cutoff of the $\delta$ distribution on $X$, i.e. modes  
$|n|>\frac{M-1}{2}$ are discarded. 

Given a continuum function 
$F\in L$ we call $f_M=I_M^\ast\cdot F\in l_M$ or $F_M=P_M\cdot F\in L_M$ the discretisation of 
$F$ at resolution 
$M$. In particular, if we have a Hamiltonian field theory on $X$ with conjugate pair of fields
$(\Phi,\Pi)$ i.e. the non-vanishing Poisson brackets are 
\be \label{a.16}
\{\Pi(x),\Phi(y)\}=\delta_X(x,y)=\sum_{n\in \mathbb{Z}}\; e_n(x-y)
\ee
then their discretisations obey
\be \label{a.16a}
\{\pi_{M,m},\phi_{M,\hat{m}}\}=M\;\delta_{m,\hat{m}},\;\;
\{\Pi_M(x),\Phi_M(y)\}=P_M(x,y)=\frac{\sin(M\pi(x-y))}{\sin(\pi(x-y))}
\ee
The latter formula is known as the Dirichlet kernel.

Given a functional $H[\Pi,\Phi]$ of the continuum fields we define its discretisation 
by 
\be \label{a.16b}
h_M[\pi_M,\phi_M]:=H_M[\Pi_M,\Phi_M]=H[\Pi_M,\Phi_M]=(I'_M H)[\pi_M,\phi_M]
\ee
where $I'_M$ denotes the pull-back by $I_M$. That is,
in the continuum formula for $H$ one substitutes $\Pi\to \Pi_M,\; \Phi\to \Phi_M$
in the formula for $H$ 
upon which $H$ is restricted to $\Pi_M,\Phi_M$, i.e. $H_M$ is that restriction, 
and then uses the identity $\Pi_M=I_M\cdot \pi_M,\;\Phi_M=I_M\cdot \phi_M$. In order for this 
to be well-defined it is important that $I_M$ is sufficiently smooth as $H$ typically 
depends of derivatives of $\Pi,\Phi$. This is granted by our choice of $I_M$. In particular,
as the derivative $\partial=\frac{\partial}{\partial x}$ preserves each of the spaces $L_M$ 
we have a canonical discretisation of the derivative defined by 
\be \label{a.17}
\partial_M:=I_M^\ast \cdot \partial\cdot I_M
\ee
which obeys $\partial_M^n=I_M^\ast \cdot \partial^n\cdot I_M$ because 
$I_M\cdot I_M^\ast=P_M$ and $[\partial,P_M]=0$. 

Concerning quantisation, in the the continuum we define the Weyl algebra $\mathfrak{A}$ 
generated by the Weyl elements 
\be \label{a.18}
W[F]=e^{-i<F,\Phi>_L},\; W[G]=e^{-i<G,\Pi>_L}
\ee 
for real valued $F,G\in L$ (or a dense subspace thereof with additional properties 
such as smoothness and rapid momentum decrease of its Fourier modes 
$<e_n,F>_L, \; <e_n,G>_L$). That is, the non-trivial Weyl relations are  
\ba \label{a.19}
&& W[G]\; W[F]\; W[-G]=e^{-i<G,F>_L}\;W[F],\; 
W[F]\;W[F']=W[F+F'], \; W[G]\;W[G']=W[G+G']
\nonumber\\
&& W[0]=1_{\mathfrak{A}},\;
W[F]^\ast=W[-F],\; W[G]^\ast=W[-G] 
\ea
Cyclic representations $(\rho,{\cal H},\Omega)$ of $\mathfrak{A}$ with $\Omega\in {\cal H}$ 
a cyclic vector (i.e. ${\cal D}:=\rho(\mathfrak{A}){\cal H}$ is dense) are generated from states 
(positive, linear, normalised functionals)
$\omega$ on $\mathfrak{A}$ via the GNS construction \cite{p}. The correspondence is given 
by 
\be \label{a.20}
\omega(A)=<\Omega, \; \rho(A)\Omega>_{{\cal H}}  
\ee

We may proceed analogously with the discretised objects. For each $M$ we define 
the Weyl algebra $\mathfrak{A}_M$ generated by the Weyl elements  
\be \label{a.21}
W_M[F_M]=e^{-i<F_M,\Phi_M>_{L_M}}=w_M[f_M]=e^{-i<f_M,\phi_M>_{l_M}},\;\;
W_M[G_M]=e^{-i<G_M,\Pi_M>_{L_M}}=w_M[g_M]=e^{-i<g_M,\pi_M>_{l_M}}
\ee
where $F_M=I_M \cdot f_M,\; G_M=I_M\cdot g_M$ are real valued. Accordingly
\ba \label{a.22}
&&W_M[G_M]\; W_M[F_M]\; W_M[-G_M]=e^{-i<G_M,F_M>_{L_M}}\;W_M[F_M],\; 
W_M[F_M]\; W_M[F'_M]=W_M[F_M+F'_M],\;
\\
&& 
W_M[G_M]\; W_M[G'_M]=W_M[G_M+G'_M],\;
W_M[0]=1_{\mathfrak{A}_M},\;W_M[F_M]^\ast=W_M[-F_M],\; W_M[G_M]^\ast=W_M[-G_M] 
\nonumber
\ea 
and completely analogous for $\phi_M,\pi_M$ if we substitute lower case letters for capital letters
in (\ref{a.22}).
For each $M$ we define a state $\omega_M$ on $\mathfrak{A}_M$ which gives rise 
to GNS data $(\rho_M, {\cal H}_M, \Omega_M)$ and the dense subspace ${\cal D}_M=
\mathfrak{A}_M \Omega_M$. 
Note that $\mathfrak{A}_M$ is a subalgebra of $\mathfrak{A}_{M'}$ for $M<M'$ and 
that $\mathfrak{A}_M$ is a subalgebra of $\mathfrak{A}$. This follows from the 
identities 
\be \label{a.23}
W_{M'}[F_M]=W_M[F_M],\;\;W_M[F_M]=W[F_M]
\ee
due to $P_{M'} \cdot P_M= P_M$ since $L_M\subset L_{M'}$ and $P_M\cdot P_M=P_M$ respectively.
 
The sole reason for discretisation is as follows: While finding states on $\mathfrak{A}$ 
is not difficult (e.g. Fock states) it is tremendously difficult to find such states 
which allow to define non-linear functionals of $\Pi,\Phi$ such as Hamiltonians 
densely on $\cal D$ due to UV singularities arising from the fact that $\Pi,\Phi$ 
are promoted to operator valued distributions whose product is a priori ill-defined.
In the presence of the UV cut-off $M$ this problem can be solved because e.g. 
$\Phi_M(x)^2$ is perfectly well-defined ($\Phi$ is smeared with the smooth kernel 
$P_M$). Suppose then that $h_M$ or equivalently $H_M$ are somehow quantised
on ${\cal D}_M$. We denote these quantisations by 
$\rho_M(h_M,c_M)$ or $\rho_M(H_M,c_M)$ respectively to emphasise that these operators are 
1. densely defined on $\rho_M(\mathfrak{A}_M)\Omega_M$, 2. correspond to the classical symbol 
$h_M$ of $H_M$ respectively and 3. depend on a set of choices $c_M$ for each $M$ 
such as  factor or normal ordering etc. It is therefore not at all clear 
whether the theories defined for each $M$ in fact descend from a continuum theory. 
By ``descendance'' we mean that $\omega_M$ is the restriction of $\omega$ to 
$\mathfrak{A}_M$ and that $\rho_M(H_M,c_M)$ is the restriction of $\rho(H,c)$ to ${\cal D}_M$
as a quadratic form (i.e. in the sense of matrix elements). In formulas this means 
\begin{align} \label{a.24}
\omega_M(A_M)&=\omega(A_M), \\
<\rho_M(A_M)\Omega_M,\;\rho_M(H_M,c_M)\; \rho_M(B_M)\Omega_M>_{{\cal H}_M}
&=<\rho(A_M)\Omega,\;\rho(H,c)\; \rho(B_M)\Omega>_{{\cal H}}, \nonumber
\end{align}
for all $M\in \mathbb{O}$ and all $A_M,B_M\in\mathfrak{A}_M$. If they did,
then we obtain the following identities for $M<M'$
\begin{align} 
\label{a.25}
\omega_{M'}(A_M) & =\omega_M(A_M),\\
<\rho_M(A_M)\Omega_M,\;\rho_M(H_M,c_M)\; \rho_M(B_M)\Omega_M>_{{\cal H}_M}
& = <\rho_{M'}(A_M)\Omega_{M'},\;\rho_{M'}(H_{M'},c_{M'})\; \rho_{M'}(B_M)\Omega_{M'}>_{{\cal H}_{M'}}, \nonumber
\end{align}

called consistency conditions. This follows from the fact that $A_{M'}:=A_M,
B_{M'}:=B_M$ can be considered as elements of $\mathfrak{A}_M$ and then using (\ref{a.24}).
With some additional work \cite{d} one can show that (\ref{a.25}) are necessary and sufficient
for $\omega, \rho(H)$ to exist (at least as a quadratic form).

In constructive quantum field theory (CQFT) \cite{b} one proceeds as follows. One starts with 
an Ansatz of a family of discretised theories
$(\omega^{(0)}_M,\rho^{(0)}_M(H_M,c^{(0)}_M))_{M\in \mathbb{O}}$. That Ansatz generically 
violates (\ref{a.25}). We now define a renormalisation flow of states and quantisations 
by defining the sequence   
$(\omega^{(k)}_M,\rho^{(k)}_M(H_M,c^{(k)}_M))_{M\in \mathbb{O}}$ for $k\in \mathbb{N}_0$ via 
\ba\label{a.26}
&& \omega^{(k+1)}_M(A_M):=\omega^{(n)}_{M'(M)}(A_M),\;\;
<\rho_M^{(k+1)}(A_M)\Omega^{(k+1)}_M,\;\rho^{(k+1)}_M(H_M,c^{(k+1)}_M)\; \rho^{(k+1)}_M(B_M)\Omega^{(k+1)}_M>_{{\cal H}^{(k+1)}_M}
\nonumber\\
&=& <\rho^{(k)}_{M'}(A_M)\Omega^{(k)}_{M'},\;\rho^{(k)}_{M'}(H_{M'},c^{(k)}_{M'})\; \rho^{(k)}_{M'}(B_M)\Omega^{(k)}_{M'}>_{{\cal H}^{(k)}_{M'}}
\ea
where $M':\mathbb{O}\to \mathbb{O}$ is a fixed map with the property that 
$M'(M)>M,\;M'(M)\not=M$. The first relation defines a new state at the coarser 
resolution $M$ as the restriction of the old state at the finer resolution $M'(M)$. 
This then defines also new GNS data $(\rho^{(k+1)}_M,{\cal H}^{(k+1)}_M,\Omega^{(k+1)}_M)$ via the GNS construction.
The second relation defines the matrix elements of an operator or quadratic form
in that new representation 
and with new quantisation choices to be made at coarser 
resolution 
in terms of the restriction of the matrix elements of the old operator or quadratic form with old 
quantisation choices in the old representation at finer resolution. A fixed point 
family $(\omega^\ast_M,\rho^\ast_M(H_M,c^\ast_M))_{M\in \mathbb{O}}$
of the flow (\ref{a.26}) solves (\ref{a.25}) at least for $M'=M'(M)$ and all $M$
and thus all $[M']^n(M),\;n\in \mathbb{N}_0$ and all $M$. This typically implies 
that (\ref{a.25}) holds for all $M'<M$. In practice we will work with $M'(M):=3
\;M$  
  
Note that for a general operator or quadratic form $O$ defined densely on $\cal D$
it is not true that we find an element $a\in \mathfrak{A}$ such that $\rho(a)=O$
(e.g. unbounded operators) which is why the above statements cannot be made just in terms of 
the states $\omega$. If one tried, one would need to use sequences or nets 
$a_n\in \mathfrak{A}$ whose
limits lie outside of $\mathfrak{A}$. On the other hand, if one prefers to work with the Weyl elements
$W_M[F_M]$ one may relate the spaces $l_M, \; L_M$ via the identities 
$w_M[f_M]=W_M[F_M],\; f_M=I_M^\ast\cdot F_M$. The $w_M[f_M], \; w_{M'}[f_M']$ at resolution $M, \; M'$ 
respectively can be related via the \textit{coarse graining} map $I_{M M'}:= I_{M'}^\ast\cdot I_M;\; l_M\to l_{M'}$ such that
$w_{M'}[I_{M M'}\cdot f_M]=w_M[f_M]$. This map obeys $I_{M_2 M_3}\cdot I_{M_1 M_2}=I_{M_1 M_3}$ for 
$M_1<M_2<M_3$ because the image of $I_M$ is $L_M$ which is a subspace of $L_{M'}$ thus
$I_{M_2 M_3}\cdot I_{M_1 M_2}=I_{M_3}^\ast\cdot P_{M_2}\cdot  I_{M_1}=I_{M_3}^\ast\cdot I_{M_1}$. 
Then  $W_{M'}[F_M]=w_{M'}[I_{M'}^\ast F_M]=w_{M'}[I_{M'}^\ast\cdot I_M\cdot f_M]
=w_{M'}[I_{M M'}\cdot f_M]$ indeed. For the same reason $W_{M'}[F_M]=W_M[F_M]$ as 
$L_M$ is embedded in $L_{M'}$ by the identity map. The renormalization flow in terms of Weyl elements 
$w_M[f_M]$ and the coarse graining map $I_{M,M'}$ takes the form
\begin{align}
\label{rflow}
\omega_M^{(k+1)}(w_M[f_M])&:=\omega^{(k)}_{M'}(w_{M'}[I_{M,M'}f_M']) \nonumber \\ 
\braket{w_M[f'_M]\Omega_{M}^{(k+1)},H^{(k+1)}_{M}\;w_M[f_M]\Omega_{M}^{(k+1)}}_{\mathcal{H}^{(k+1)}_M}&:=
\braket{w_{M'}[I_{M,M'}f'_M]\Omega_{M'}^{(k)},H^{(k)}_{M'}\;w_{M'}[I_{M,M'}f_M]\Omega_{M'}^{(k)}}_{\mathcal{H}^{(k)}_{M'}}.
\end{align}

\end{appendix}


\begin{thebibliography}{99}

\parskip -5pt

\bibitem{a} R. Haag, ``Local Quantum Physics'', Springer Verlag, Berlin,
1984

\bibitem{b} J. Glimm and A. Jaffe, ``Quantum Physics'',
Springer Verlag, New York, 1987.
   
\bibitem{c} K. G. Wilson. The renormalization group:
Critical phenomena and the Kondo
problem. Rev. Mod. Phys. {\bf 47} (1975) 773

\bibitem{d} T. Thiemann. Canonical quantum gravity, constructive QFT and
renormalisation.
Front. in Phys. \textbf{ 8} (2020) 548232. arXiv:2003.13622 [gr-qc].

\bibitem{LLT1} T. Lang, K. Liegener, T. Thiemann.
 Hamiltonian Renormalisation I.
Derivation from Osterwalder-Schrader Reconstruction.
Class. Quant. Grav. {\bf 35} (2018) 245011.
[arXiv:1711.05685]

\bibitem{LLT2} T. Lang, K. Liegener, T. Thiemann. Hamiltonian Renormalisation II.
Renormalisation Flow of 1+1 dimensional free, scalar fields: Derivation.
Class. Quant. Grav. {\bf 35} (2018) 245012.
[arXiv:1711.06727]

\bibitem{LLT3} T. Lang, K. Liegener, T. Thiemann. Hamiltonian Renormalisation III.
Renormalisation Flow of 1+1 dimensional free, scalar fields: Properties.
Class. Quant. Grav. {\bf 35} (2018) 245013.
[arXiv:1711.05688]

\bibitem{LLT4} T. Lang, K. Liegener, T. Thiemann. Hamiltonian Renormalisation IV. Renormalisation Flow of D+1 dimensional
free scalar fields and Rotation Invariance.
Class. Quant. Grav. {\bf 35} (2018) 245014, [arXiv:1711.05695]

\bibitem{LT} K. Liegener, T. Thiemann.
Hamiltonian Renormalisation V. Free Vector Bosons.
Front. Astron. Space Sci. {\bf 7} (2021) 547550.
e-Print: 2003.13059 [gr-qc]

\bibitem{TT} T. Thiemann.
Hamiltonian Renormalisation VII. Free Fermions and doubler free kernels.
Phys. Rev. {\bf D108} (2023) 12, 125007.
e-Print: 2207.08291 [hep-th]

\bibitem{TZ} E.-A. Zwicknagel. Hamiltonian renormalization. VI. Parametrized field theory on the cylinder.
Phys. Rev. {\bf D108} (2023) 12, 125006. e-Print: 2207.08290 [gr-qc]

\bibitem{RZ-T} 
M. Rodriguez Zarate, T. Thiemann. Hamiltonian renormalisation VIII. $P(\Phi)_2$ quantum field theory

\bibitem{e} B. Simon, ``The P($\phi$)2 Euclidean (Quantum) Field Theory'',
Princeton Unviersity Press, 1974

\bibitem{f} L. Smolin. The G(Newton) $\to 0$
limit of Euclidean quantum gravity. Class. Quant. Grav. {\bf 9} (1992)
883-894. e-Print: hep-th/9202076 [hep-th]

\bibitem{g} T. Thiemann.
Exact quantisation of U(1)$^3$ quantum gravity via exponentiation of the hypersurface deformation algebroid.
Class. Quant. Grav. {\bf 40} (2023) 24, 245003. e-Print: 2207.08302 [gr-qc]

\bibitem{h} H. Narnhofer, W.E. Thirring.
Covariant QED without indefinite metric.
Rev. Math. Phys. {\bf 4} (1992) spec01, 197-211

\bibitem{i} T. Thiemann. Nonperturbative quantum gravity in Fock representations.
Phys. Rev. {\bf D 110} (2024) 12, 124023. e-Print: 2405.01212 [gr-qc]

\bibitem{j} C. Rovelli, ``Quantum Gravity'', Cambridge University
Press, Cambridge, 2004.\\
T. Thiemann, ``Modern Canonical Quantum General Relativity'', Cambridge
University Press, Cambridge, 2007.\\
J. Pullin, R. Gambini, ``A first course in Loop Quantum Gravity'',
Oxford University Press, New York, 2011\\
C. Rovelli, F. Vidotto, ``Covariant Loop Quantum Gravity'', Cambridge
University Press, Cambridge, 2015\\
K. Giesel, H. Sahlmann,
From Classical To Quantum Gravity: Introduction to Loop Quantum Gravity,
PoS QGQGS2011 (2011) 002, [arXiv:1203.2733].

\bibitem{k} J. Glimm, ``Boson Fields with the $:\phi^4:$ Interaction in Three
Dimensions", Comm. Math. Phys. {\bf 10} (1968)  1-47.\\
J. Glimm, A. Jaffe, Positivity of the $\phi^4_3$ Hamiltonian'',
Fortschr. Phys. {\bf 21} (1973) 327–376

\bibitem{m}  I. Daubechies. Ten lectures of wavelets. Springer Verlag, Berlin,
1993.\\
A. Cohen, I. Daubechies, P.Vial. Wavelets on the interval and
fast wavelet transforms. Appl. and Comp. Harm. Anlysis, Elsevier, 1993.
[hal-01311753]

\bibitem{l} T. Thiemann. Renormalization, wavelets, and the Dirichlet-Shannon kernels. Phys. Rev. D {\bf 108} 
(2023) 12, 125008. e-Print: 2207.08294 [hep-th]

\bibitem{n} S. Bakhoda, T. Thiemann.
Covariant origin of the $U(1)^3$ model for Euclidean quantum gravity
Class. Quant. Grav. {\bf 39} (2022) 2, 025006. e-Print: 2011.00031 [gr-qc].\\
S. Bakhoda, T. Thiemann. Reduced Phase Space Approach to the $U(1)^3$ model for Euclidean
Quantum Gravity. Class. Quantum Grav. {\bf 38} (2021) 215006.
e-Print: 2010.16351 [gr-qc]

\bibitem{o} S. Bakhoda, T. Thiemann.
Asymptotically Flat Boundary Conditions for the
$U(1)^3$ Model for Euclidean Quantum Gravity. Universe 7 (2021) 3, 68.
e-Print: 2010.16359 [gr-qc]\\
S. Bakhoda, Y. Ma.
Geometrical quantum time in the U(1)$^3$ model of Euclidean quantum gravity.
Commun. Theor. Phys. {\bf 77} (2025) 5, 055401. e-Print: 2411.19435 [gr-qc]

\bibitem{p} O. Bratteli, D. W. Robinson, ``Operator Algebras and Quantum
Statistical
Mechanics'', vol. 1,2, Springer Verlag, Berlin, 1997.

\bibitem{q} T. Thiemann.
Non-degenerate metrics, hypersurface deformation algebra, non-anomalous representations and density weights in quantum gravity.
Gen. Rel. Grav. {\bf 56} (2024) 10, 122. e-Print: 2207.08299 [gr-qc]
K. Giesel, T. Thiemann. Hamiltonian Theory: Dynamics. In {\it Handbook of quantum gravity},
C. Bambi, L. Modesto, I. Shapiro (eds.), Springer Verlag, Berlin, 2024. 
e-Print: 2303.18172 [gr-qc]

\bibitem{r} T. Thiemann.
Quantum Spin Dynamics (QSD) : V.
Quantum Gravity as the Natural Regulator of the Hamiltonian Constraint
of Matter Quantum Field Theories.
Class. Quantum Grav. {\bf 15} (1998) 1281-1314, [gr-qc/9705019]

\bibitem{s} T. Thiemann. Quantum Spin Dynamics (QSD).
Class. Quantum Grav. {\bf 15} (1998) 839-73, [gr-qc/9606089];
Quantum Spin Dynamics (QSD): II.
The Kernel of the Wheeler-DeWitt Constraint Operator.
Class. Quantum Grav. {\bf 15} (1998) 875-905, [gr-qc/9606090];
Quantum Spin Dynamics (QSD): III. Quantum constraint algebra and physical scalar product in quantum general relativity
Class. Quant. Grav. {\bf 15} (1998) 1207-1247. e-Print: gr-qc/9705017 [gr-qc]

\bibitem{t} K. Giesel, T. Thiemann, Scalar Material Reference Systems and Loop Quantum
Gravity, Class. Quant. Grav. \textbf{32} (2015) 135015,
[arXiv:1206.3807 [gr-qc]]. 

\bibitem{u} T. Thiemann. Quantum spin dynamics. VIII. The Master constraint.
Class. Quant. Grav. {\bf 23} (2006) 2249-2266, [gr-qc/0510011]

\bibitem{v} R. Gambini, J. Lewandowski, D. Marolf, J. Pullin
On the consistency of the constraint algebra in spin network quantum gravity.
Int. J. Mod. Phys. {\bf D 7} (1998) 97-109; e-Print: gr-qc/9710018 [gr-qc]

\bibitem{v1} A. Laddha.
Hamiltonian constraint in Euclidean LQG revisited:
First hints of off-shell Closure. e-Print: 1401.0931 [gr-qc]\\
M. Varadarajan. Anomaly free quantum dynamics for Euclidean LQG.
e-Print: 2205.10779 [gr-qc]


\bibitem{Klausthesis}
K. Liegener.
\textit{Renormalisation in Loop Quantum Gravity}.
Ph.D. thesis, Friedrich-Alexander-Universität Erlangen-Nürnberg, 2019.
\end{thebibliography}
\end{document}